\documentclass[acmsmall,nonacm,screen]{acmart}

\acmJournal{JACM}

\usepackage{amsmath,amsfonts}
\usepackage{algorithmic}
\usepackage{graphicx}
\usepackage{textcomp}
\usepackage{xcolor}

\usepackage{multirow}

\usepackage{fancyhdr}
\usepackage[normalem]{ulem}
\usepackage{array}
\usepackage{subfig}
\usepackage{flushend}
\usepackage{balance}
\usepackage{verbatim}
\usepackage{breqn}
\usepackage{enumitem}
\usepackage{color}
\usepackage{soul}
\soulregister\cite7
\soulregister\ref7
\soulregister\pageref7
\usepackage{wrapfig}

\newcommand{\edge}[2]{#1$\rightarrow$#2}

\begin{document}

\title{Calipers: A Criticality-aware Framework for Modeling Processor Performance}
\author{Hossein Golestani}
\affiliation{%
  \institution{University of Michigan}
  \city{Ann Arbor}
  \state{MI}
  \country{USA}}
\email{hosseing@umich.edu}
\author{Rathijit Sen}
\affiliation{%
  \institution{Microsoft}
  \city{Madison}
  \state{WI}
  \country{USA}}
\email{rathijit.sen@microsoft.com}
\author{Vinson Young}
\affiliation{%
  \institution{Microsoft}
  \city{Redmond}
  \state{WA}
  \country{USA}}
\email{vinson.young@microsoft.com}
\author{Gagan Gupta}
\affiliation{%
  \institution{Microsoft}
  \city{Redmond}
  \state{WA}
  \country{USA}}
\email{gagg@microsoft.com}

\begin{abstract}
Computer architecture design space is vast and complex.
Tools are needed to explore new ideas and gain insights quickly, with low efforts and at a desired accuracy.
We propose Calipers, a criticality-based framework to model key abstractions of complex architectures and a program's execution using \textit{dynamic event-dependence graphs}.
By applying graph algorithms, Calipers can track instruction and event dependencies, compute critical paths, and analyze architecture bottlenecks.
By manipulating the graph, Calipers enables architects to investigate a wide range of Instruction Set Architecture (ISA) and microarchitecture design choices/``what-if'' scenarios during both early- and late-stage design space exploration \textit{without recompiling and rerunning the program}.
Calipers can model in-order and out-of-order microarchitectures, structural hazards, and different types of ISAs, and can evaluate multiple ideas in a single run.
Modeling algorithms are described in detail.

We apply Calipers to explore and gain insights in complex microarchitectural and ISA ideas for RISC and EDGE processors, at lower effort than cycle-accurate simulators and with comparable accuracy.
For example, among a variety of investigations presented in the paper, experiments show that targeting only a fraction of critical loads can help realize most benefits of value prediction.
\end{abstract}

\maketitle

\section{Introduction}

Computer architects use a variety of performance analysis tools to evaluate techniques and optimize modern processor designs.
Different tools have different capabilities and provide different levels of insights in the design~\cite{6813138}. 
For example, an Instruction-Set Simulator (ISS) can prove functional correctness and is fast, but can only count the total instructions executed.
Cycle-Accurate Simulators (CAS) can provide more accurate execution time (in cycles), but can be much slower~\cite{binkert:gem5:can:2011,sniper,wenisch2005simflex,yourst2007ptlsim,sanchez2013zsim,patel2011marss}. 
FPGAs can emulate the design even more accurately, but can require almost as much effort as designing the hardware~\cite{firesim}.
Mechanistic models~\cite{ino_mechanistic, ooo_mechanistic, highlevel_mechanistic} may also be used to analyze performance, but require manually building analytical penalty models. 
Often, Cycles-Per-Instruction (CPI) stack~\cite{perf_cpi} is built using these tools to count where cycles were spent to execute an instruction. 
CPI stacks can reveal sources of lost cycles, but only in the aggregate over the program.

Although useful to study performance, these tools do not provide direct insights into program critical paths needed when optimizing designs, as is also noted by others~\cite{multistage_cpi,fields:interaction-costs:micro:2003}. 
Furthermore, exploring the design space using such tools requires modeling or implementing the design choices.
This can require changing the compiler or modifying the simulator/FPGA, which can be non-trivial. 
Moreover, simulations may have to be rerun, which can take long, to the tune of even weeks.

To obtain insights related to program critical paths, prior works have proposed criticality-aware tools that model program execution using dependence graphs and perform microarchitectural bottleneck analysis~\cite{tdg_techreport,robatmili:expoiting-criticality:hpca:2011,fields:interaction-costs:micro:2003,nagarajan:critical-path-trips:ispass:2006}. 
Such analysis can reveal the precise instruction sequences and processor resources in the critical paths, which can become targets for optimization.
These proposals rely on a CAS for cycle estimates to compute the critical paths.
Most proposals also rely on the CAS to evaluate new design choices, while others also use graph transformation~\cite{tdg_techreport}.
We extend this line of work by proposing a more versatile and capable framework, {\em Calipers}.

Calipers treats criticality as a first class citizen. 
It constructs a \textit{Dynamic Event-dependence Graph} (DEG) of a program's execution on a given architecture. 
DEG, a DAG, captures three key aspects: events during an instruction's execution, their sequence of occurrence, and the interval between them. 
Events correspond to microarchitectural operations, e.g., instruction fetch, memory access, execute, commit, etc.
The sequence denotes ordering between events, which can arise from data dependencies, structural hazards, etc.
Intervals capture time, e.g., in cycles, between the events. The program's run time is then just the length of the critical path in the DEG. 

Once constructed, besides using the DEG to identify performance bottlenecks, Calipers allows architects to manipulate the DEG to model different design points simultaneously and evaluate them using graph analysis without requiring program reruns.
We make the following contributions:

\setlist{nolistsep}
\begin{itemize}[noitemsep, leftmargin=*]
\item \textbf{Modeling}: We demonstrate how Calipers uses a unified dependence-graph--based framework to model a variety of ISA and microarchitecture designs at the desired fidelity.
    \begin{itemize}
    \item We model In-Order (InO) as well as Out-of-Order (OoO) execution, conventional as well as dataflow architectures, pipelined resources, branch speculation, and value prediction. 
    Prior dependence-graph proposals do not tackle such a diverse range of architectures and scenarios (Section~\ref{sec:related}).
    \item We model structural hazards and instruction scheduling, not described in prior work, to enable accurate analysis, particularly when exploring design choices. We present the modeling algorithms in detail and examine their space and time complexities in comparison to alternatives.
    \item While prior work uses CAS~\cite{fields:interaction-costs:micro:2003,robatmili:expoiting-criticality:hpca:2011,nagarajan:critical-path-trips:ispass:2006} or a hardware shotgun profiler~\cite{fields:interaction-costs:micro:2003} for event costs to build the dependence graph, Calipers also supports statistical and feature-specific models to permit analysis upon unavailability of detailed models like CAS, mechanistic models or real hardware, e.g., during early design exploration.
    \end{itemize}
\item \textbf{Rapid Exploration}: We introduce \textit{Vectorized Graph Analysis}, to linearly speed up design exploration. By representing design parameters as vectors, multiple choices can be evaluated simultaneously at the desired accuracy, e.g., we evaluated 32 configurations in a single run at a speedup of $\sim$14$\times$.
\item \textbf{Insights}: 
We show that Calipers can model performance as accurately as conventional simulators while also providing insights in performance bottlenecks, critical paths, and design choices that may be difficult to obtain from other tools such as the ISS, CAS, CPI stacks, etc. For example:
    \begin{itemize}
    \item For RISC-V~\cite{Waterman:risc5-isa:EECS-2014-54} processors, bottleneck and limit studies reveal that optimizing the branch predictor might suffice for an in-order core, but an out-of-order core may also benefit from optimizing the instruction fetch. Furthermore, a criticality-aware value prediction scheme can realize most of the speed-up opportunity by accurately predicting $\sim$15\% of the loads. 
    \item For EDGE processors~\cite{burger:edge-architecture:computer:2004}, EDGE-specific instructions, \texttt{READ} and \texttt{MOV}, introduce non-negligible overheads, but \texttt{NULL} does not. We also evaluate three alternatives for instruction block formats, without needing a new compiler, and identify that for best performance, circuit optimizations may also be needed.
    \end{itemize}    
\item \textbf{Open-source Release}:
To support further research on these topics, we plan to open-source our implementation of Calipers. To the best of our knowledge, this would be the first open-source dependence-graph--based tool for critical path analysis.
\end{itemize}

We start by making a case for a graph-framework approach like Calipers in Section~\ref{sec:case}.
Calipers details are then described in Section~\ref{sec:graph-framework}. 
We present accuracy validation and performance results for Calipers in Section~\ref{sec:accuracy-validation}. 
Next, we present case studies using Calipers to analyze bottlenecks and explore design choices for RISC-V in Section~\ref{sec:risc5-model} and EDGE architecture in Section~\ref{sec:edge-model}. We discuss related work in Section~\ref{sec:related}, before concluding in Section~\ref{sec:conclusion}.

\section{A Case for Graph-based Performance Modeling}
\label{sec:case}

\begin{figure}[t]
\centering
\includegraphics[width=0.48\textwidth]{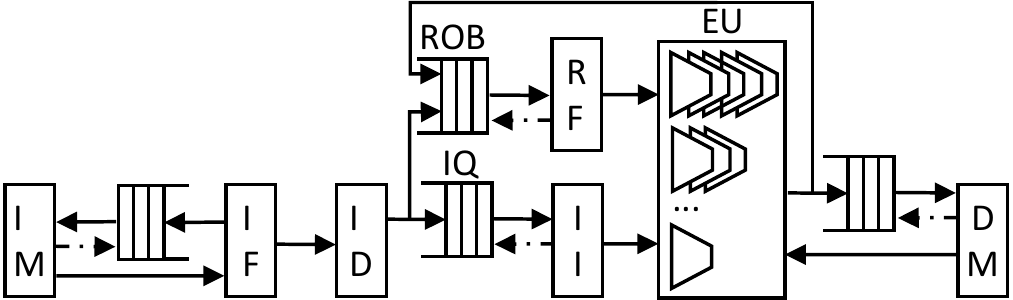}
\caption{An abstract view of a high-performance processor. IM: instruction memory; IF: instruction fetch; ID: instruction decode; IQ: instruction queue; ROB: reorder buffer; RF: register file; II: instruction issue; EU: execution units; DM: data memory. Dotted arrows imply back-pressure.}
\label{fig:pipeline}
\vspace{-8pt}
\end{figure}

A processor's performance is a function of how the microarchitecture that implements the ISA processes the program.
A modern processor executes a program using a logical pipeline of functional units (Figure~\ref{fig:pipeline}).
An instruction-fetch unit (IF) fetches instructions from the instruction memory (IM), which are issued (II) for execution to multi-stage execution units (EUs).
During execution, instructions may fetch data from the register file (RF) or the data memory (DM).
Computed results may be committed to the RF or the DM.
A high-performance implementation may execute instructions in order or out of order, and the functional units may operate concurrently while communicating with each other, and exploit Instruction-Level Parallelism (ILP) by processing multiple instructions simultaneously.

A program's flow through this pipeline is frequently interrupted by three key dynamic hazards (dependencies): data, control and structural.
Instructions cannot execute until data dependencies are resolved, or be fetched until control dependencies are resolved, or proceed until resources are available.
These hazards, a function of the microarchitecture and the program characteristic, can diminish the ILP or stall the flow intermittently.
To handle interruptions to the flow, inter-stage queues may be used to hold instructions until they can proceed.
Thus the IF unit dispatches instructions (ID) to the instruction queue (IQ) from where they are issued (II), possibly out-of-order, to execution units.
Results are queued in the reorder buffer (ROB) until they can be committed in order.
Memory requests may also be buffered in queues.

Concurrent OoO operations, queuing effects, dynamically varying parallelism, and pipeline stalls drastically complicate reasoning about performance.
We simplify such analysis by mapping an abstract view of the ISA, the microarchitecture, the program flow, and the effects of the hazards to a graphical domain and applying graph analysis.

\begin{figure}[t]
\centering
\includegraphics[width=1\textwidth]{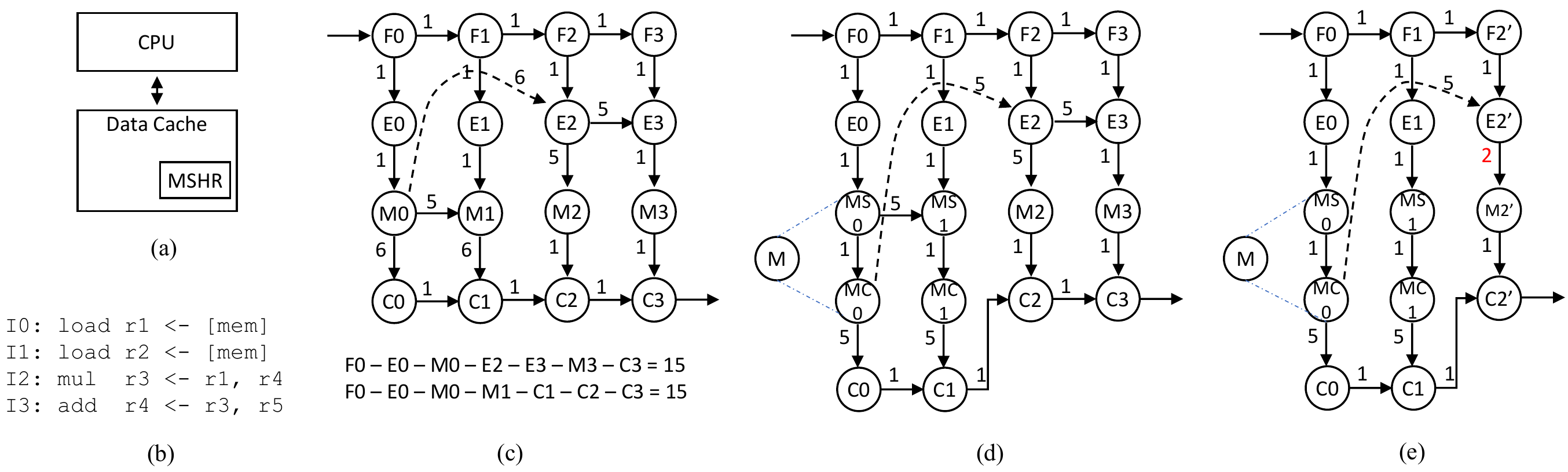}
\vspace{-24pt}
\caption{Modeling execution using event graphs. (a) Example system. (b) Example program. (c) Event graph of the execution. (d) Modeling event M at higher fidelity. (e) Exploring design space by manipulating the graph.}
\label{fig:case}
\vspace{-12pt}
\end{figure}

Consider a simple CPU and data cache (D\$) system shown in Figure~\ref{fig:case}(a). 
Say the CPU implements a simple 4-stage pipeline---Fetch (F), Execute (E), Memory (M), and Commit (C).
Say the D\$ returns data to a \texttt{load} instruction within a cycle on a hit.
On a miss, it takes a cycle to register the request in a Miss Status Holding Register (MSHR) and another 5 cycles to fetch data from the memory.

Consider the program in Figure~\ref{fig:case}(b). 
I3 has data dependence on I2, which in turn is dependent on I0. 
Say both I0 and I1 \texttt{load}s miss in the D\$.
The \texttt{mul} (I2) takes 5 cycles to produce the result.
A simulator can simulate and report that the program takes 15 cycles to execute on the example system.
However, it provides very little insights into where each instruction spends time, the program's critical path, bottlenecks, etc., which are needed to optimize the system.
Intuitively, it might seem that a faster multiplier would speed up the program, but one cannot be sure until the new design is implemented and simulated again, which can take long.
Alternatively, one may compute the CPI stack, a single figure of merit that gives the breakdown of where an instruction spends cycles on average over the whole program.
Although useful to study a design at hand, it again gives no insight in which specific aspect should be optimized first (more in Section~\ref{sec:what-if-fetch-bp}).

\textbf{Bottleneck Analysis.}
Modeling system events as a graph and tracking criticality can provide such insights more easily.
Figure~\ref{fig:case}(c) shows the execution modeled as an {\em event graph}.
Each vertex denotes an event during the instruction's execution---Fetch (F), Execute (E), Memory (M), and Commit (C)---and the edges represent dependencies and ordering constraints.
For example, the edge \edge{M0}{E2} denotes data dependence, and the edge \edge{C0}{C1} represents in-order commit.
Each edge has a ``cost'', represented by its weight, denoting cycles taken to perform the action.
The longest path in this graph and its composition gives insights in the critical path, e.g., the multiplier is in the critical path (F0-E0-M0-E2-E3-M3-C3) and can be a target for optimization.
However, graph analysis also reveals that the longest non-multiplier path (F0-E0-M0-M1-C1-C2-C3) is also 15 cycles long.
Therefore, speeding up just the multiplier will not help.
Such insights are not easy to obtain from a simulator or a CPI stack.

\textbf{What-if Exploration.}
Furthermore, a closer scrutiny can be applied by selectively modeling the graph in additional detail.
The M vertex can be expanded to two vertices: MS and MC, representing logging of the D\$ request in the MSHR and retrieving data from the memory, respectively (Figure~\ref{fig:case}(d)).
This reveals the MSHR to be a structural hazard---the second request is blocked until the first vacates it, denoted by the \edge{MS0}{MS1} edge with a weight of 5.
Therefore, optimizing for this program requires eliminating this hazard by introducing a second MSHR.
Removing the \edge{MS0}{MS1} edge and reducing the cost of E2 vertex from say 5 to 2 cycles, representing a faster multiplier, in the graph can reduce the critical path to 12 cycles (Figure~\ref{fig:case}(e)), a 20\% improvement.
Such changes can be evaluated by simply manipulating and analyzing the graph.
It is non-trivial to evaluate such tweaks using a simulator or a CPI stack.

\textbf{Early-stage ISA Exploration.}
Event graphs also simplify exploring other types of ideas, such as accelerators, e.g., combining \texttt{mul} (I2) and \texttt{add} (I3) instructions into a 2-cycle \texttt{mac}.
Using a simulator to evaluate this ISA enhancement would require changes in the compiler to produce new code and in the simulator to model the new instruction. 
In a criticality-aware graph-based approach, \texttt{mul} followed by \texttt{add} can be modeled as combined vertices (F2', E2', M2', C2'), and reanalyzing the graph for critical paths shows a speedup of 27\% (Figure~\ref{fig:case}(e)).

Calipers enables such a graph-based approach to analyze bottlenecks, evaluate performance, and explore the design space by manipulating the graph at a desired fidelity.

\section{Calipers}
\label{sec:graph-framework}
Prior proposals have presented dependence graphs comprising event vertices and dependency edges derived from CAS or hardware runs (Section~\ref{sec:related}).
We extend this concept by introducing \textit{vector-weighted graphs}, formally defining them, and taking inputs from multiple sources to build the graph (Section~\ref{sec:framework-overview}).
Additionally, we model data/control speculation (Section~\ref{sec:spec}), structural hazards (Section~\ref{sec:issue-and-hazards}), microarchitecture features such as instruction fusion/cracking, pipelined execution units (Section~\ref{sec:misc}), two types of ISA (Section~\ref{sec:edge}), multiple configurations through a vectorized graph (Section~\ref{sec:vec}), and InO and OoO instruction scheduling (Section~\ref{sec:alg}).
\subsection{Calipers Architecture}
\label{sec:framework-overview}

To describe Calipers, we start with a generic model of the ISA and the microarchitecture, and then extend it to variants such as OoO pipelines and the EDGE ISA.

To model dynamic program execution, we introduce the notion of a \textit{vector-weighted} directed acyclic \textit{Dynamic Event-dependence Graph} (DEG), $G=(V,E,W)$, comprising a set of vertices, $V$, a set of edges, $E$, and a set of vector of weights, $W$.
 A vertex $v\in V$ denotes the microarchitectural event a dynamic instance of an instruction undergoes, e.g., Fetch, Issue, Execute, Commit, etc., depending on the 
 modeled microarchitecture.
 $E=\{(v_s,v_d,\hat{w})|v_s,v_d\in V,\hat{w}\in W,|\hat{w}|\geq 0, \hat{w} = \{w_0,w_1,\ldots, w_n\}, w_i\geq 0\}$ 
 is the set of edges connecting the vertices in $G$, denoting dependence between source vertex, $v_s$, and destination vertex, $v_d$, and
 $\hat{w}$ denotes a vector of $n$ weights associated with the edge.
 Assume scalar edge weights, $w$, in the following discussion until we visit vectorized graph analysis, without loss of generality.

The directed edge models an \textit{ordering dependency} between a pair of events, and the edge weight models the delay, or \textit{minimum latency lag}, between them. Thus, the event corresponding to $v_d$ can only start $w$ cycles (or later) after $v_s$; it can start only when the lags for all incoming edges to $v_d$ are satisfied. Ordering arises from essential microarchitectural characteristics (e.g., fetch before dispatch), or from the three types of hazards.
A weighted directed path in the graph denotes the sequence of events and the associated delays.
Disjoint paths denote concurrent events.
Together, paths in the DEG characterize a program's execution and temporal properties. 

\begin{figure}[t]
\centering
\includegraphics[width=0.6\textwidth]{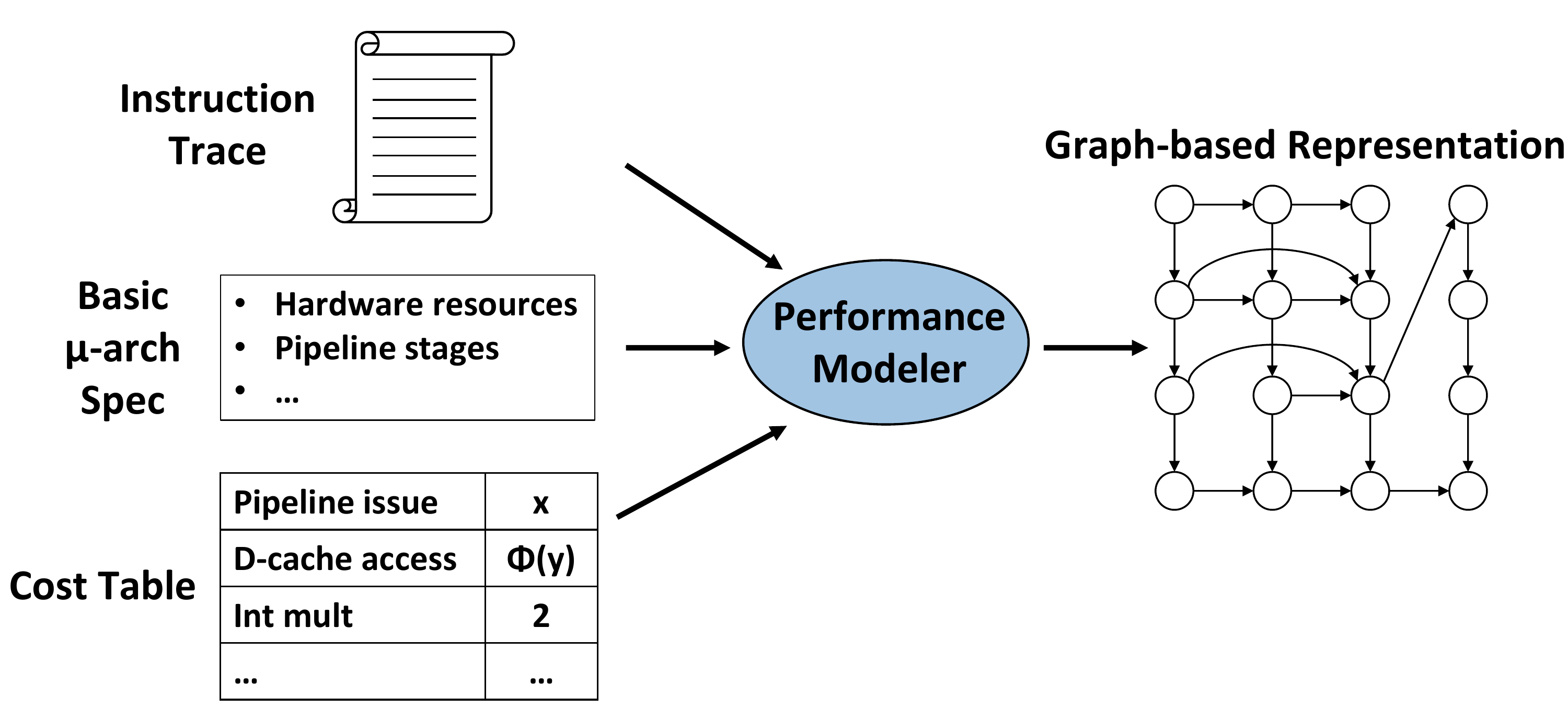}
\vspace{-5pt}
\caption{High-level overview of Calipers graph-based model construction.}
\label{fig:overview-interface}
\vspace{-20pt}
\end{figure}

Calipers takes the program's dynamic instruction trace, microarchitectural and ISA specifications of the processor, and a \textit{cost table} as inputs from the user (Figure~\ref{fig:overview-interface}).
It builds the DEG by identifying vertices from the dynamic microarchitectural events. 
Edges for data and control dependencies are discovered from the instruction trace.
Resource dependencies are modeled using the microarchitectural details---such as issue order (in-order or out-of-order), structural resources (functional units, load/store queues, etc.), and their bandwidths. We describe different vertex types and details of modeling different types of dependencies later in this section.

Graph edge weights are provided by the cost table. 
They represent microarchitectural latencies in cycles. Some costs are deterministic, such as pipeline inter-stage latencies and latency of execution units (e.g., adder, multiplier, floating-point unit). 
However, latencies of instruction and data memory accesses may be non-deterministic due to hierarchical cache/memory architecture. 
Latency of individual loads/stores and instruction cache line accesses may be captured and provided along with the trace, e.g., using a CAS or even real hardware.
However, often \textit{CAS and real hardware are unavailable} during early design exploration.
In such cases, users may obtain instruction traces from an ISS and costs from statistical models  or functional/analytical cache simulators.  
A similar approach may be used to determine branch prediction results (correct or incorrect).
We demonstrate these different options in Sections~\ref{sec:accuracy-validation}-\ref{sec:edge-model}.

Once the DEG is constructed, performance and bottlenecks can be analyzed by applying graph algorithms (e.g., longest-path search) to compute total execution cycles. 
Further, Calipers facilitates efficient design exploration using edge-weight vectors, wherein multiple configurations or design scenarios can be evaluated in a single pass of graph analysis (Section~\ref{sec:vec}). 
By analyzing the length and composition of the critical path of the DEG, Calipers can identify ILP and design bottlenecks. 
Analyzing secondary critical paths can help understand to what extent a specific component in the primary critical path is worth optimizing. 
Furthermore, by adjusting weights of graph edges or even manipulating graph vertices, various what-if scenarios can be modeled and explored without rerunning the program. 
We evaluate real-life what-if scenarios that architects may face in their everyday practice using Calipers (Sections~\ref{sec:risc5-model} and \ref{sec:edge-model}).

\begin{table}[t]
    \centering
	\footnotesize
	\caption{Basic vertices and edges, and modeling novelty achieved by expanding or manipulating them.}
\vspace{-10pt}
	\label{tab:basic-edges}
	\begin{tabular}{|>{\arraybackslash}m{1.7cm}|>{\arraybackslash}m{5.5cm}|>{\arraybackslash}m{5.5cm}|} \hline
    	\textbf{Edge}& \textbf{Weight}& \textbf{Calipers Novelty}\\ \hline \hline
    	$F_n \rightarrow E_n$& Cycles of decoding/dispatching instruction $n$& Flexible pipeline front-end modeling, e.g., different decode/dispatch and issue bandwidths (Section~\ref{sec:misc}); Modeling various instruction encoding formats in the EDGE ISA (Section~\ref{sec:edge-blocks})\\ \hline
    	$E_n \rightarrow C_n$& Execution cycles of instruction $n$& Flexible pipeline back-end modeling, e.g., pipelined functional units and instruction cracking/fusing (Section~\ref{sec:misc})\\ \hline
    	$F_n \rightarrow F_{n+1}$& 0 (in-order fetch), or cycles of predicting and fetching the cache line containing instruction $n+1$ (in case of correct prediction)& \multirow{2}{5.5cm}[5pt]{Modeling block-structured ISAs such as EDGE by combining $F$ vertices (Sections~\ref{sec:misc} and \ref{sec:edge-model}); Branch predictor and I-cache interface optimization study (Section~\ref{sec:what-if-fetch-bp})}\\ \cline{1-2}
    	$F_n \rightarrow F_{n+fbw}$& 1, where $fbw$ is fetch bandwidth in terms of number of instructions&\\ \hline
    	$E_n \rightarrow F_{n+1}$& Cycles of executing/resolving the mispredicted branch (instruction $n$), and fetching the cache line containing instruction $n+1$& Modeling control misspeculation (Section~\ref{sec:modeling}); Branch predictor and I-cache interface optimization study (Section~\ref{sec:what-if-fetch-bp})\\ \hline
    	$E_n \rightarrow E_m$& Execution cycles of instruction $n$ or cycles required to forward execution result of instruction $n$ to instruction $m$ (register/memory data dependency)& Modeling structural hazards (Section~\ref{sec:issue-and-hazards}); Modeling data speculation (Sections~\ref{sec:modeling} and \ref{sec:value-pred})\\ \hline
    	$C_n \rightarrow C_{n+1}$& 0 (in-order commit)& \multirow{2}{5.5cm}[-5pt]{Modeling block-structured ISAs such as EDGE by combining $C$ vertices (Section~\ref{sec:misc} and \ref{sec:edge-model})}\\ \cline{1-2}
    	$C_n \rightarrow C_{n+cbw}$& 1, where $cbw$ is commit bandwidth in terms of number of instructions&\\ \hline
    \end{tabular}
\vspace{-15pt}
\end{table}

\textbf{Modeling.}
\label{sec:modeling}
Calipers builds and analyzes a dependence graph using various vertex and edge types similar to prior work~\cite{fields:interaction-costs:micro:2003, nagarajan:critical-path-trips:ispass:2006,tdg_techreport,robatmili:expoiting-criticality:hpca:2011}.
Table~\ref{tab:basic-edges} summarizes the basic edge types and event costs for modeling pipeline, data, and control dependencies. 
Figure~\ref{fig:case} illustrates a sample DEG for the example system and program. 
Next, we elaborate on modeling details novel to Calipers.
These include new types or uses of vertices and edges to model microarchitecture features.
After analyzing the DEG, new features may be explored to overcome shortcomings. 
When exploring choices, such as different number, types, and performance of various execution units/components, the DEG may have to be reconstructed and reanalyzed. 
We present algorithms to do so.

\subsection{Speculation}
\label{sec:spec}
\begin{figure}[t]
\centering
\includegraphics[width=0.70\textwidth]{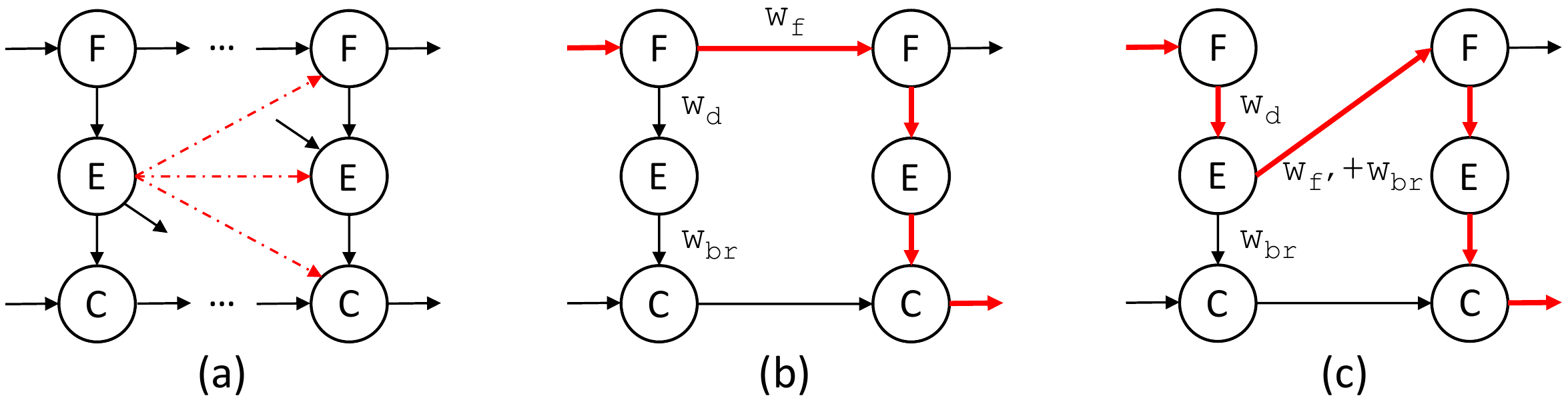}
\vspace{-10pt}
\caption{(a) Handling misspeculation in general; red dotted edge(s) may be added from the speculated to the dependent instruction to model misspeculation penalty. (b) Correct branch prediction. (c) Incorrect branch prediction. (Bold red edges show critical paths. $w_d$ and $w_{br}$: decode/dispatch and execution cycles of the branch; $w_f$ and $w_{f'}$: fetch cycles of the next instruction)}
\label{fig:speculation}
\vspace{-13pt}
\end{figure}

Control speculation (branch prediction) and data speculation (value prediction~\cite{lipasti1996value}) in general are modeled by appropriately inserting and/or removing edges between the speculated and the dependent instructions, as shown in Figure~\ref{fig:speculation}(a).
Once misspeculation is detected, typically by checking the speculated instruction's result, a few cycles may be spent freeing up resources---such as queues, execution units, and the re-order buffer---before the execution can resume.
This recovery is modeled by adding/modifying edge(s) between the E vertex of the speculated instruction and the affected vertices of the dependent instruction.

Calipers models data misspeculation by adding penalty or cleanup cycles to the weight of the associated data dependency edge (i.e., \edge{E$_n$}{E$_m$} in Table~\ref{tab:basic-edges}). In contrast, correct value prediction is simply modeled by removing the edge \edge{E$_n$}{E$_m$}.

Figures~\ref{fig:speculation}(b) and \ref{fig:speculation}(c) show modeling of correct and incorrect control speculation, respectively.
Note that graph vertices correspond to the correct program execution path regardless of branch prediction correctness.
Assuming instruction $n$ is a branch, Calipers inserts the edge \edge{F$_n$}{F$_{n+1}$} for correct prediction; otherwise, it inserts the edge \edge{E$_n$}{F$_{n+1}$}.
The weight of the edge \edge{F$_n$}{F$_{n+1}$} is the cycles of predicting and fetching instruction $n+1$, whereas the weight of the edge \edge{E$_n$}{F$_{n+1}$} comprises execution cycles of the branch, branch misprediction penalty (if any), and cycles of fetching instruction $n+1$.
Using \edge{E$_n$}{F$_{n+1}$} instead of \edge{F$_n$}{F$_{n+1}$} may lengthen the critical path, e.g., Figure~\ref{fig:speculation} shows that the critical path in the case of misprediction is $w_d+w_{br}+(w_{f'}-w_f)$ cycles longer than the correct prediction. 
Note that $w_{f'}$ is often larger than $w_f$, as an I-cache miss following branch misprediction is quite probable.

\subsection{Structural Hazards and Event Order}
\label{sec:issue-and-hazards}
Finite resources in a processor lead to structural hazards due to contention, which Calipers models using the instruction trace and details of structural resources---e.g., issue bandwidth, number of Arithmetic Logic or Floating-Point Units (ALUs/FPUs), their latencies, Load/Store Queue (LSQ) size, etc. 
Events that occur out of program order, e.g., OoO instruction issue and execution, further complicate the modeling.
We discuss these two aspects using issue bandwidth and instruction execution scheduling on limited functional units as examples.
Detailed InO and OoO modeling and scheduling algorithms are described in Section~\ref{sec:alg}.

To model limited issue bandwidth and execution schedule, Calipers needs to obtain an issue order for instructions, i.e., order for E vertices. 

\textbf{In-order Core}. In-order issue is simply modeled by inserting an edge between the E vertices of consecutive instructions in program order, i.e., \edge{E$_n$}{E$_{n+1}$}, with the weight of zero. Consequently, limited issue bandwidth can be modeled by \edge{E$_n$}{E$_{n+ibw}$} edges with the weight of one, where $ibw$ is the issue bandwidth. If an \edge{E$_n$}{E$_{n+1}$} or \edge{E$_n$}{E$_{n+ibw}$} edge already exists due to data dependency, it is unnecessary to add it again.

\textbf{Out-of-order Core}. Calipers models instruction issue and use of structural resources in an OoO core using critical path analysis.
Calipers first builds the base graph for a sliding window of instructions, accounting for control and data hazards, and calculates the length of the critical path to E vertices.
Instructions are issued and scheduled to use resources using the list of E vertices sorted by their critical path length (see Section~\ref{sec:alg} for more details).
For modeling limited issue bandwidth, say the obtained order is E$_a^1$, E$_b^2$, E$_c^3$, ... (superscripts denote issue order; subscripts denote program order; the two are not necessarily the same). 
Then, \edge{E$_x^n$}{E$_y^{n+ibw}$} edges with the weight of 1 impose limited issue bandwidth of $ibw$.

Based on the obtained issue order, Calipers adds edges to model limited resources.
It maintains a scoreboard with entries for each resource to track when the resource will be available next (Section~\ref{sec:alg}).
Say there are $m$ FPUs, and $i_1$, $i_2$, $i_3$, ... is the prioritized list of floating-point instructions ($f_1$, $f_2$, $f_3$, ...). \edge{E$_{f_x}^{i_n}$}{E$_{f_y}^{i_{n+m}}$} edges model the order in which these instructions use the FPUs, which is equivalent to Least Recently Used (LRU) scheduling. The edge weight denotes the number of cycles after which $f_y$ can use the FPU that $f_x$ is using (i.e., 1 if pipelined or latency if non-pipelined).

\begin{figure}[t]
\centering
\includegraphics[width=0.70\textwidth]{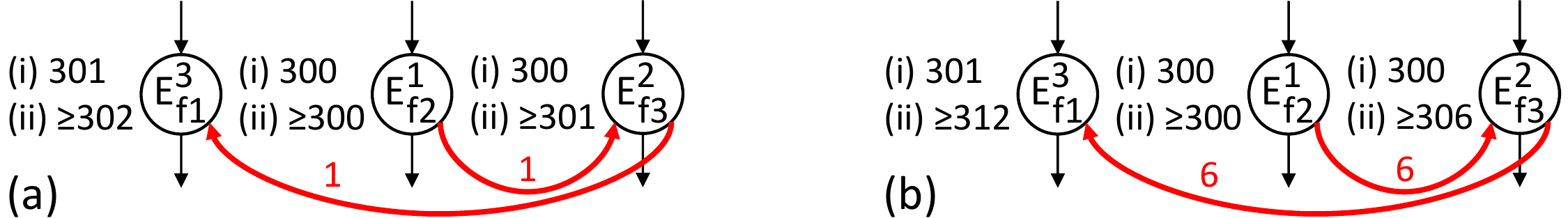}
\caption{Modeling structural hazards of contention for a single FPU by three floating-point instructions when the FPU is: (a) pipelined, or (b) not pipelined. FPU operation latency is 6 cycles. Numbers next to each vertex show the related critical path length to it (i) before, and (ii) after adding structural-hazard--related edges (red/bold edges).}
\vspace{-15pt}
\label{fig:structural-hazard}
\end{figure}

Adding structural edges can cause critical path lengths of children vertices to change, which Calipers accounts for in subsequent processing, as illustrated in Figure~\ref{fig:structural-hazard}.
Assume that three floating-point instructions contend for a single FPU. Numbers marked ``(i)'' show the length of the critical path to each E vertex prior to adding the structural-hazard edges, which is the basis for scheduling. 
Figures~\ref{fig:structural-hazard}(a) and~\ref{fig:structural-hazard}(b) show the cases for pipelined and non-pipelined FPUs, respectively. 
Since $f_2$ is scheduled first, adding \edge{E}{E} edges to model the contention for the FPU lengthens the critical path to $f_1$ and $f_3$ in both cases. 
Numbers marked ``(ii)'' show a lower bound for critical path length because adding structural-hazard--related edges corresponding to vertices preceding these three vertices in the graph 
could also lengthen the critical path.
The algorithms for obtaining the scheduling order, adding structural hazard edges, and updating the critical path are described in Section~\ref{sec:alg}.

\subsection{Modeling Additional Microarchitecture Features}
\label{sec:misc}

Calipers models additional microarchitecture features using variations of the fundamentals presented so far.

\begin{figure}[t]
\centering
\includegraphics[width=0.60\textwidth]{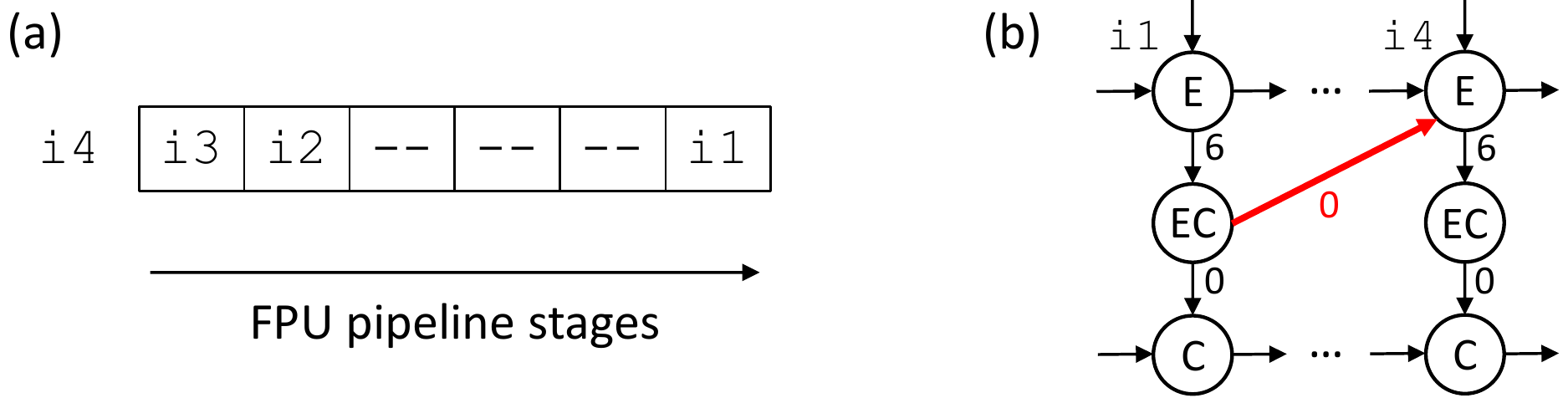}
\vspace{-12pt}
\caption{Limitation in the pipeline of a sample FPU: (a) Timing snapshot in real execution, (b) Modeling the limited capacity of the FPU with the red (bold) edge in the graph.}
\label{fig:fpu-limit-example}
\vspace{-12pt}
\end{figure}

\textbf{New Vertex Types}. More vertex types can be introduced for modeling more details or just for more flexibility. We provide three examples.
First, a decode/dispatch vertex (D) may be used for instructions, wherein pipeline dependencies of each instruction are modeled by F$\rightarrow$D$\rightarrow$E$\rightarrow$C edges. 
This can be useful to model differing dispatch and issue bandwidths. 
Second, for execution of loads and stores, Calipers may use another vertex type, M, indicating that the memory request is ready to be sent, to decouple modeling of address calculation and memory request, which increases modeling flexibility. 
As such, pipeline dependencies of a load/store would be F$\rightarrow$E$\rightarrow$M$\rightarrow$C. 
Third, besides limited number of functional units (Section~\ref{sec:issue-and-hazards}), limited number of pipeline stages in a functional unit can also introduce structural hazards. 
This is important in modeling cases where an instruction is stalled in one of the pipeline stages. 
Figure~\ref{fig:fpu-limit-example}(a) is a real-case snapshot of four floating-point instructions $i_1$-$i_4$ (not necessarily in program order) scheduled to use an FPU with six pipeline stages. 
$i_4$ can only use the FPU after $i_1$ completes because of the intermediate pipeline bubbles. 
Calipers models this scenario (Figure~\ref{fig:fpu-limit-example}(b)) with a new vertex type, Execution Completed (EC). 
Note that we cannot simply model this case with the \edge{C$_{i_1}$}{E$_{i_4}$} edge, particularly in an OoO model. If $i_4$ precedes $i_1$ in program order, then the \edge{C$_{i_1}$}{E$_{i_4}$} (FPU pipeline limit), \edge{E$_{i_4}$}{C$_{i_4}$}, and \edge{C$_{i_4}$}{C$_{i_1}$} (in-order commit) edges create a loop in the graph, which must be a DAG.

\begin{figure}[t]
\centering
\includegraphics[width=0.75\textwidth]{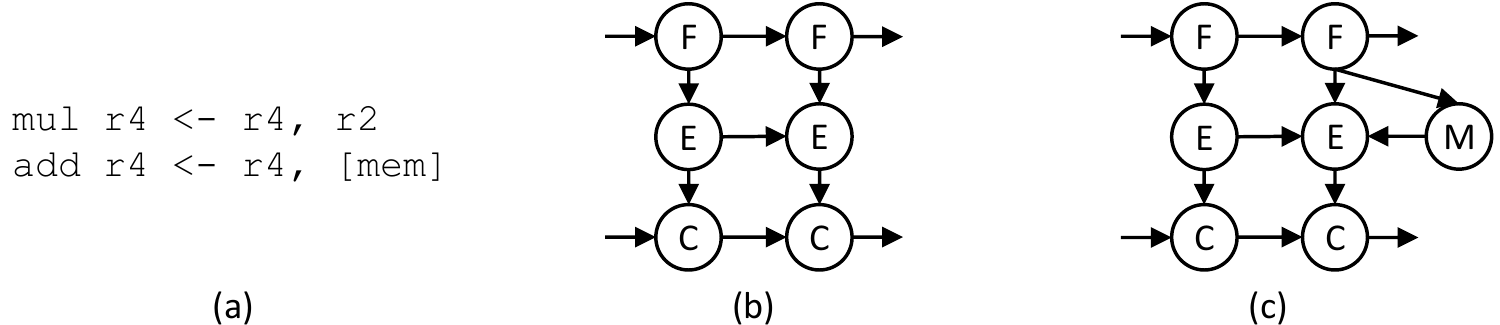}
\vspace{-10pt}
\caption{Modeling instruction cracking. (a) Example code sequence. \texttt{add} is a complex instruction.  (b) DEG subgraph for the example. (c) Vertex M is added when \texttt{add} is cracked.}
\label{fig:uop_cracking}
\vspace{-15pt}
\end{figure}

\textbf{Instruction Cracking}. 
CISC ISAs can have complex instructions which are decomposed (\textit{cracked}) into simpler \textit{micro-ops} to increase ILP opportunities~\cite{pentiumpro}.
This feature can be modeled by introducing additional vertices and edges for the micro-ops.
Figure~\ref{fig:uop_cracking}(a) shows an example code sequence of \texttt{mul} that updates register \texttt{r4}, followed by a complex \texttt{add} instruction that adds the value from memory location \texttt{mem} to \texttt{r4}.
This code sequence is modeled in Figure~\ref{fig:uop_cracking}(b).
A processor may crack the \texttt{add} into a memory and an add operations.
The memory operation can now execute concurrently with the preceding \texttt{mul}.
This is modeled by adding the M vertex and edges to the subgraph of the code sequence (Figure~\ref{fig:uop_cracking}(c)), so that Calipers can also account for the ILP.

\textbf{Instruction Fusion}.
Conversely, a processor may also fuse instructions into a \textit{macro-op}~\cite{Gochman2003TheIP}. 
This can be modeled by collapsing the corresponding vertices into a single vertex while retaining the edges (as seen in Figure~\ref{fig:case}(e)).

\vspace{-5pt}
\subsection{Modeling EDGE ISA}
\label{sec:edge}
Besides RISC and CISC, Calipers is generic enough to model other ISAs, e.g., the emerging Explicit Data Graph Execution (EDGE) ISA.
We highlight key aspects of EDGE relevant to our work and refer readers for details to related literature~\cite{burger:edge-architecture:computer:2004, clp, robatmili:expoiting-criticality:hpca:2011}.

EDGE programs are composed from statically demarcated \textit{blocks} of instructions, analogous to if-converted hyperblocks~\cite{hyperblocks}.
The program's execution advances by blocks.
Within a block, instructions can communicate results using broadcast channels or directly to consumers by encoding their target IDs.
\texttt{MOV} instruction is provided to fanout results when consumers are many.
\texttt{READ} instruction is provided to communicate General-Purpose Register (GPR) values to instructions.
A block may consist of if-converted predicated code.
Within a block, number of writes to GPRs and memory in both taken and not-taken predicated paths are required to be balanced.
\texttt{NULL} instruction is provided to simulate dummy GPR/memory writes to balance out the paths if needed.

\begin{figure}[t]
\centering
\includegraphics[width=0.40\textwidth]{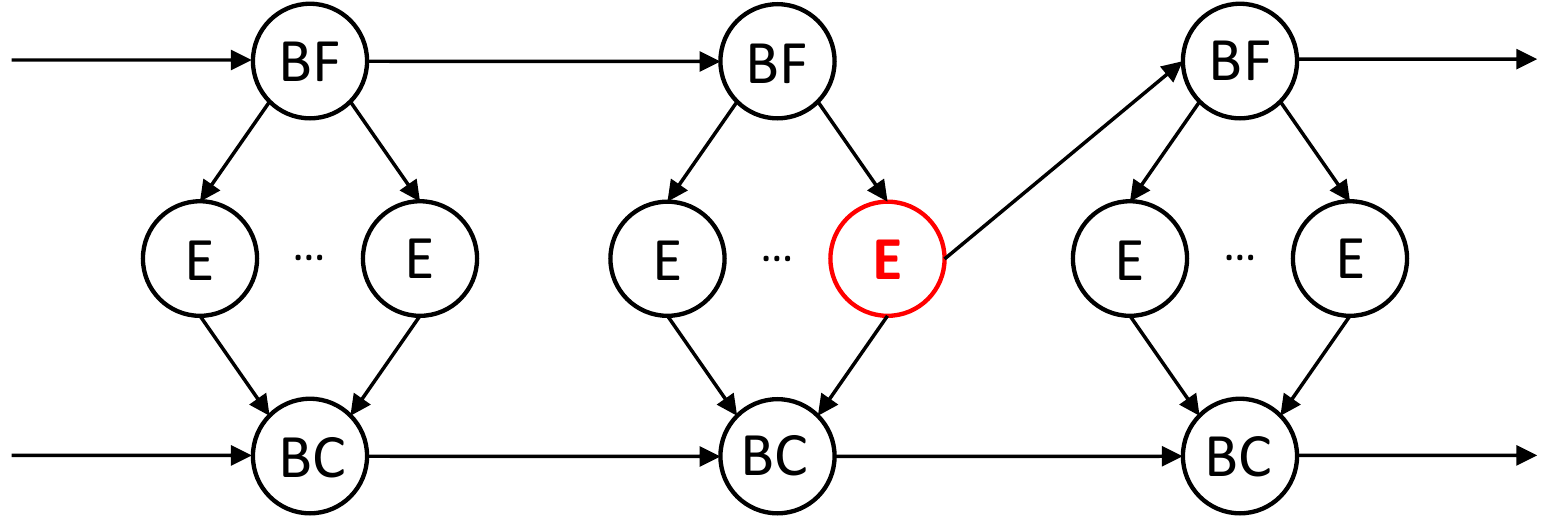}
\vspace{-5pt}
\caption{Sample graph model for three EDGE blocks. The red bold vertex corresponds to a mispredicted branch.}
\label{fig:edge-graph-example}
\vspace{-15pt}
\end{figure}

To model EDGE, Calipers maps EDGE events to the conventional vertices and edges (Table~\ref{tab:basic-edges}). It replaces the F vertex with BF (Block Fetch), to denote that the block is fetched, and the C vertex with BC (Block Commit), to denote that the block is ready to commit (Figure~\ref{fig:edge-graph-example}). 
An edge is added from the BF vertex to the E vertex of each instruction in the block, akin to instruction cracking. 
Edges are added from E vertices of instructions whose execution is necessary for committing the block, i.e., branches, stores, and GPR-write instructions, to a single BC vertex, akin to instruction fusing.
Instruction fetch cycles are reflected in the weight of \edge{BF$_{n}$}{BF$_{n+1}$}. The weight of \edge{BC$_{n}$}{BC$_{n+1}$} is determined by the cycles needed to commit stores and GPR writes of block $n$.
Branch $i$ in block $n$, if mispredicted, is modeled by the  \edge{E$_{i,n}$}{BF$_{n+1}$} edge, and by \edge{BF$_n$}{BF$_{n+1}$} if correctly predicted.

Data dependencies, arising from broadcast channels or direct communication or through the GPR/memory, are modeled by simply adding the corresponding \edge{E}{E} edges.
Modeling issue order and structural hazards is done in a similar way as described in Section~\ref{sec:issue-and-hazards}.
Intra-block and inter-block issue may be considered separately in the case of multi-block execution. Calipers accounts for this by using issue-related \edge{E}{E} edges.

\subsection{Vectorized Graph Analysis}
\label{sec:vec}

Another novel feature of Calipers is that it can simultaneously model and analyze multiple core configurations using edge-weight vectors.
When constructing the graph, Calipers can assign a vector of weights instead of a scalar value to each edge, each element corresponding to a configuration. 
For example, N different configurations (e.g., config 1: baseline, config 2: longer decode latency, config 3: 2$\times$ ALUs, etc.), can be modeled with N elements in each edge-weight vector. 
The first element corresponds to config 1, second to config 2, and so on.
Through vectorization, Calipers does not have to construct the graph from scratch in N separate runs.
This results in almost linearly scaling Calipers' speed with vector size (Section~\ref{sec:accuracy-validation}).

\subsection{Modeling and Analysis Algorithms}
\label{sec:alg}

\begin{figure}
\centering
\includegraphics[width=0.99\textwidth]{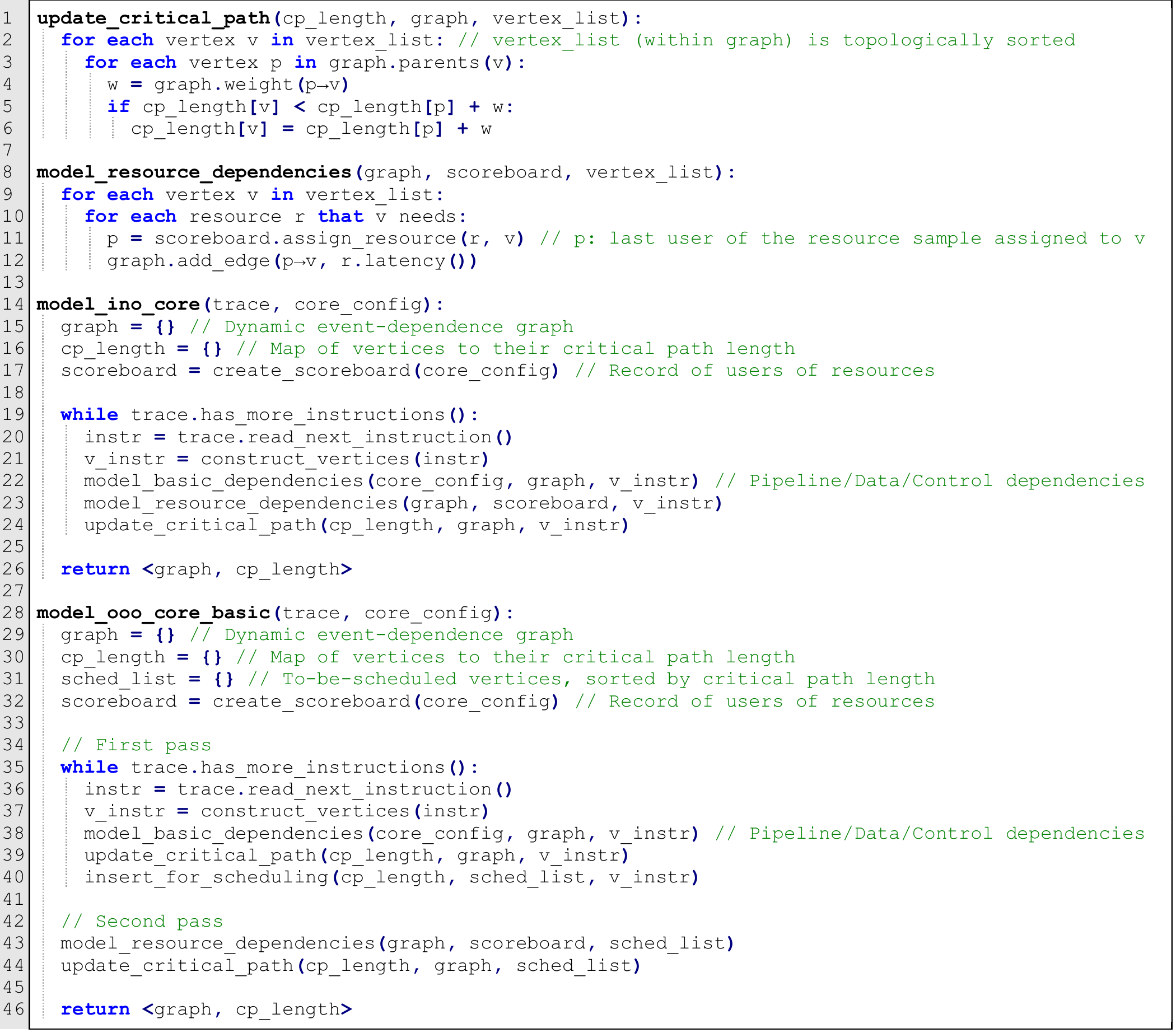}
\vspace{-10pt}
\caption{Algorithms for constructing and analyzing the DAG for the InO and OoO core models.}
\vspace{-15pt}
\label{fig:algs}
\end{figure}

DAG analysis algorithms generally consist of two steps: first, a topologically sorted list of vertices is obtained, and second, the list is traversed, e.g., to compute the longest path. 
The algorithms in Figure~\ref{fig:algs} show how the DEG is constructed, traversed, and analyzed in the context of modeling InO and OoO cores in Calipers.
In these algorithms, the \texttt{graph} data structure contains the DEG as it gets constructed, and \texttt{cp_length} maps the vertices to the length of the critical/longest path to them. The \texttt{scoreboard} data structure as well as \texttt{sched\_list} (in the OoO algorithm) are used to track and manage assignment of structural resources to vertices (Section~\ref{sec:issue-and-hazards}).

\textbf{InO Modeling.}
In the InO core model, a topologically sorted list of vertices is $V_1, V_2, V_3, ...$, wherein $V_i$ is the set of event vertices of the $i$'th instruction of the program in the pipeline order.
The InO algorithm, \texttt{model_ino_core}, reads instructions one-by-one from the trace, and builds the corresponding part of the graph (lines 20-23, Figure~\ref{fig:algs})---i.e., adds edges for modeling data, control, and resource dependencies to the vertices of the instruction. 
Length of the critical path to each vertex is calculated by considering all of its parent vertices (line 24). 
The run time of the InO algorithm is $O(|V| + |E|)$, where $|V|$ and $|E|$ are the number of vertices and edges, respectively, in the DEG. For our workloads (Section~\ref{sec:accuracy-validation}), we observed that $|E|$ grows linearly with $|V|$, which in turn is linearly proportional to the dynamic instruction count. Thus, in practice, we expect the computational complexity to be usually linear in the dynamic program size. In the case of vectorized modeling and analysis, the run time is $O(|V| + |\hat{w}|\times|E|)$ (or similarly, $O(|\hat{w}|\times|V|)$), where $|\hat{w}|$ is the length of the edge-weight vector. The space complexity of the InO algorithm depends on the \texttt{graph} and \texttt{cp_length} data structures. It is unnecessary to keep the whole DEG in memory. Only a limited ``look-back'' window of the DEG need be retained as it is being constructed, so that the information of parent vertices can be used to calculate the critical path of data-, control-, or resource-dependent vertices. As a result, the space complexity is constant. The execution time of the modeled program equals the length of the critical path to the last vertex of the last instruction.

\textbf{OoO Modeling.}
OoO scheduling in OoO cores complicates DEG construction.
In the InO case, traversing the topologically sorted list suffices since the vertices are listed in the order they are processed (i.e., program order).
This is inadequate in OoO scheduling because resource dependence edges required for modeling limited issue bandwidth, functional units, etc. may not respect the program order, as seen in Section~\ref{sec:issue-and-hazards}.
To handle this case, we use a basic two-pass algorithm, \texttt{model_ooo_core_basic} (Figure~\ref{fig:algs}).
It starts with building the base graph, which includes in-order dependencies (i.e., pipeline, data, and control), in the first pass (lines 34-40, Figure~\ref{fig:algs}). 
Vertices are inserted in the scheduling list based on the length of the critical path to them in the base graph (line 40), which, in fact, gives the priority of vertices to use structural resources. 
In the second pass (lines 42-44), resource dependence edges are added to the graph according to the scheduling list, and the length of the critical path to the resource-dependent vertices is updated.

\begin{figure}
\centering
\includegraphics[width=0.99\textwidth]{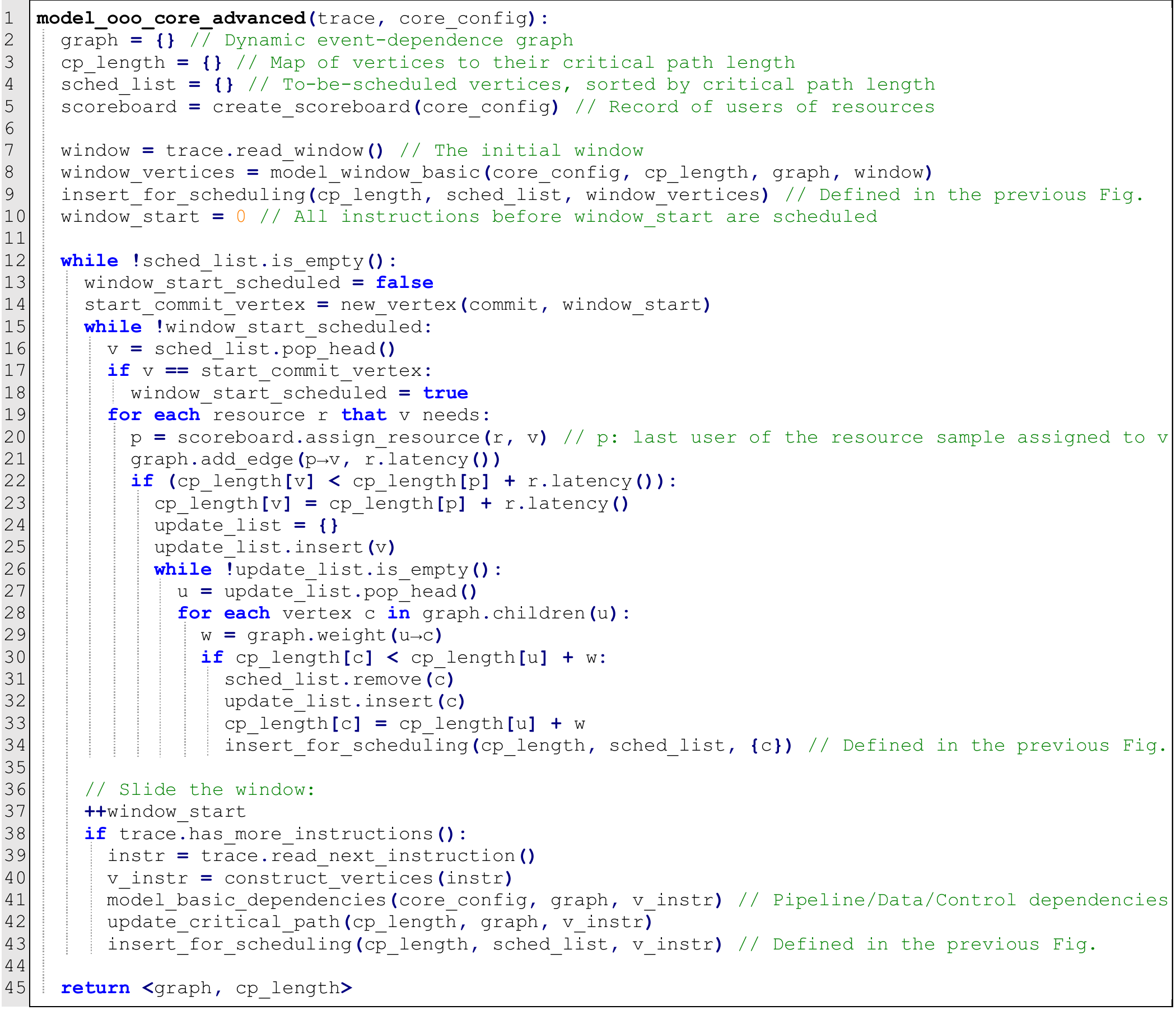}
\vspace{-5pt}
\caption{An alternative algorithm for the OoO core model with a sliding analysis window mechanism.}
\vspace{-18pt}
\label{fig:alg-ooo}
\end{figure}

The OoO algorithm in Figure~\ref{fig:algs} requires keeping the whole DEG of the modeled program in memory during model construction and analysis because the scheduling order is obtained in a separate first pass. Since the modeled program (or even subsets of it) may be arbitrarily large, the memory requirement for storing the DEG can be intractable. 
To counter this, we introduce a space-efficient, advanced algorithm, \texttt{model_ooo_core_advanced} (Figure~\ref{fig:alg-ooo}), which constructs and analyzes the DEG using sliding windows. 
It starts with building the graph for a window of instruction, i.e., the scheduling window, which includes the in-order dependencies (lines 7-10, Figure~\ref{fig:alg-ooo}). The vertices of the window are inserted to the scheduling list based on the length of the critical path to them. Resource dependence edges are added to the graph accordingly, and the length of the critical path to each resource-dependent vertex is updated (lines 12-34). Note that such an update may affect the critical path length of their dependent vertices down the scheduling window. As such, the affected dependent vertices are updated in the scheduling list (lines 24-34). Once all the vertices of the first instruction of the window are scheduled, i.e., the required resource dependence edges are added, the scheduling window is slid by one instruction, and the vertices of the new instruction are added to the scheduling list (lines 17-18 and 36-43).

Similar to the InO algorithm, the space complexity of the advanced OoO algorithm in Figure~\ref{fig:alg-ooo} is also constant because only the ``look-back'' window and the ``look-ahead'' scheduling window of the DEG need to be stored in memory. The runtime complexity of this algorithm, however, is different. Updating the critical path length of a resource-dependent vertex leads to updating its children/descendent vertices in the scheduling list (lines 12-34, Figure~\ref{fig:alg-ooo}), whose count may be on the order of the size of the scheduling window. Therefore, the runtime of the algorithm is $O(|V|\times|W|)$ (or $O(|\hat{w}|\times|V|\times|W|)$ in the case of vectorized modeling and analysis), where $|W|$ is the size of the scheduling window.
Picking a small $|W|$ can speed up the analysis, but can hurt the fidelity. We pick $|W|$ to represent the ROB size in modern OoO processors. The ROB should be large enough to capture the desired instruction-level parallelism---e.g., ROB size in Intel Skylake and Sunny Cove microarchitectures is 224 and 352, respectively~\cite{skylake, sunny_cove}. Larger ROBs result in higher hardware cost and diminishing performance improvement.

As an approximation in OoO scheduling, the active updates of the scheduling list in lines 12-34 can be disregarded (which was actually the case in the basic OoO algorithm in Figure~\ref{fig:algs}). Such an approximation may result in a different, less realistic scheduling order, but it improves the runtime to $O(|V|)$ (or $O(|\hat{w}|\times|V|)$ in the case of vectorization). We will show in Section~\ref{sec:accuracy-validation} that this approximation significantly speeds up the analysis with a slight decrease in modeling accuracy. Note that in the complexity analysis of the OoO algorithm, it is assumed that insertions of vertices to the scheduling list take O(1), as younger vertices (in program order) are likely placed at or near the end of the scheduling list; thus, efficient insertion procedures~\cite{cpp_emplace} can be leveraged.

\vspace{-5pt}
\section{Calipers Accuracy and Performance}
\label{sec:accuracy-validation}

We validated Calipers' accuracy by matching it with the gem5 simulation infrastructure~\cite{binkert:gem5:can:2011} for the RISC-V ISA~\cite{Waterman:risc5-isa:EECS-2014-54}.
We chose gem5 as the baseline for its wide usage~\cite{gem5usage}, support for different ISAs, in-order (InO) and out-of-order (OoO) cores, cycle-accurate simulation, open-source code that can be studied and modified, as well as for its accuracy~\cite{lowepower2020gem5,gutierrez2014sources}.

We matched Calipers and gem5 core models as much as possible by properly setting parameters in Calipers, such as latency of pipeline stages, fetch/issue/commit bandwidths, number of different functional units, etc. 
We modified gem5 to record I-cache/D-cache access cycles and branch resolution outcomes (correct/incorrect prediction) alongside the instruction trace.
This information was applied to model the core pipeline and data/control/resource dependencies.

\begin{figure}[t]
\centering
\includegraphics[width=0.70\linewidth]{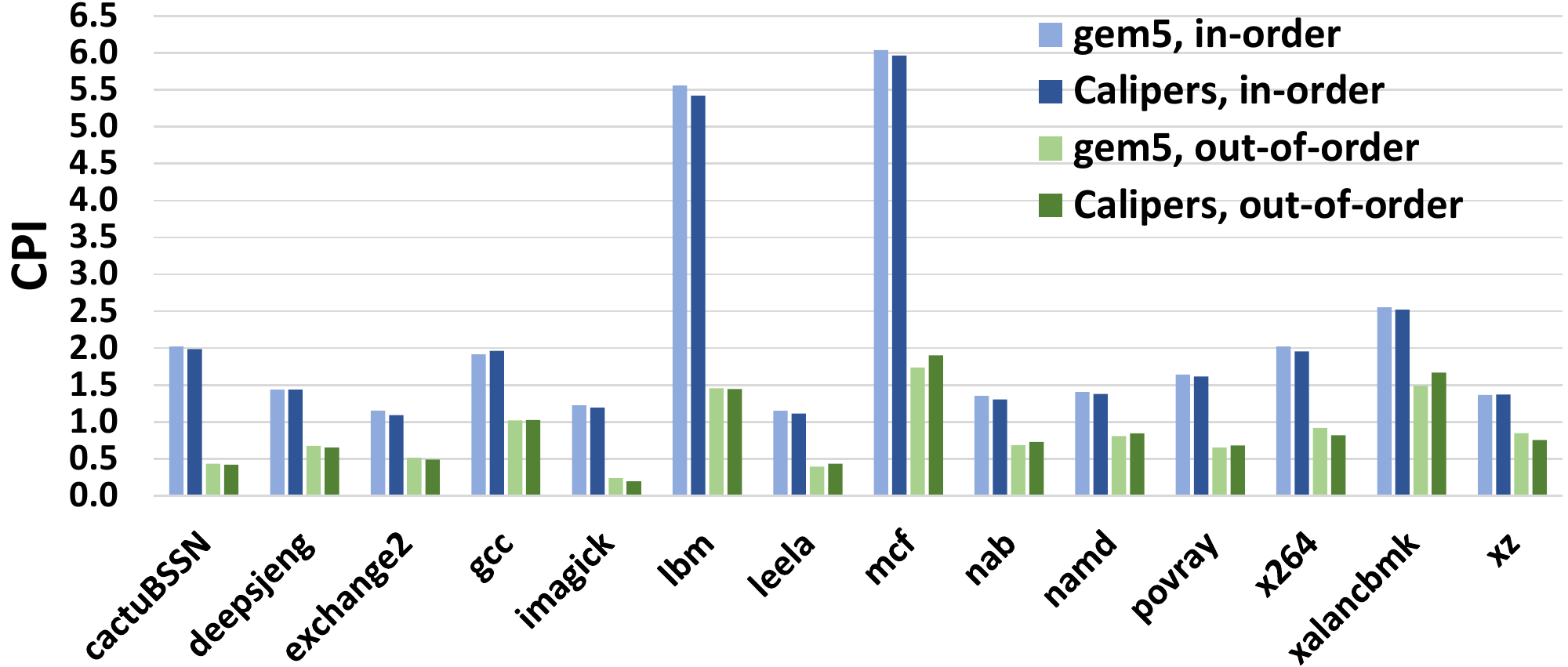}
\vspace{-10pt}
\caption{Simulated and modeled CPI for RISC-V cores.}
\vspace{-20pt}
\label{fig:risc5-accuracy-validation}
\end{figure}

\textbf{gem5 Comparison.}
We experimented with multiple configurations of gem5 InO/OoO models.
Calipers closely matched the Cycles Per Instruction (CPI) estimates and trends.
We present here results of the configuration listed in Table~\ref{tab:gem5} for the gem5 InO and OoO core models (i.e., \textit{MinorCPU} and \textit{DerivO3CPU} as called in gem5, respectively); the branch predictor and cache hierarchy are the same in both cases. gem5 simulations were performed in the System-call Emulation (SE) mode. 
We used benchmarks from the SPEC CPU 2017 suite with reference inputs. Programs are fast-forwarded by 1 billion instructions (including cache warm-up), and the next 10 million instructions are simulated. 
Whereas other subsampling approaches, e.g., SimPoints~\cite{hamerly:simpoint-3.0:jilp:2005}, are also viable, our subsamples show a wide performance range and various bottlenecks as we will see in the experimental results.
Figure~\ref{fig:risc5-accuracy-validation} shows CPI of each benchmark obtained by gem5 and Calipers. 
Geo-mean of CPI difference is 1.7\% and 4.8\% in the InO and OoO core models, respectively.

\begin{table}[t]
    \centering
    \footnotesize
    \caption{gem5 configuration.}
    \vspace{-10pt}
    \label{tab:gem5}
    \begin{tabular}{|p{1.1cm}|p{9.5cm}|} \hline
        InO core &  Pipeline stages: Fetch1, Fetch2, Decode, Execute \newline
                    2-wide issue, 2-entry LSQ \newline
                    Functional units: 2x Int-ALU, 2x Int-Mul, 2x Int-Div, 2x FPU, 2x LSU \\ \hline
        OoO core &  Pipeline stages: Fetch, Decode, Rename, IEW (Issue/Execute/Write-back), Commit \newline
                    8-wide issue, 192-entry ROB, 32-entry LQ/SQ  \newline
                    Functional units: 6x Int-ALU, 2x Int-MulDiv, 4x FP-ALU, 2x FP-MulDiv, 4x LSU \\ \hline
        Common &    Two-level tournament branch predictor \newline
                    Cache: 32 KB L1-I, 32 KB, L1-D, 1 MB LLC \\ \hline
    \end{tabular}
\vspace{-20pt}
\end{table}

The difference in CPI is generally higher in the OoO model compared to the InO model due to the OoO-specific differences between Calipers and gem5.
The precise mechanisms of OoO scheduling and attribution of misspeculation penalties are not identical.
Accuracy validation required identifying key core parameters (either configurable or not) from the multitude of parameters in gem5 and matching Calipers parameters with them. 
Additionally, recorded I-cache access cycles were carefully applied to the weights of incoming edges to F vertices. 
In both InO and OoO models of gem5, a number of I-cache accesses are related to wrong paths. We conservatively added request-to-response cycles of such accesses to the cycles of the next correct-path I-cache access when applying them to the weight of corresponding edges.

\textbf{Lower-fidelity Models.}
As discussed in Section~\ref{sec:graph-framework}, Calipers can build the DEG using different-fidelity inputs. 
We show how a lower-fidelity input enables effort vs. modeling accuracy tradeoff. Instead of using the recorded I-cache/D-cache access cycles and branch resolution outcomes from gem5, we use a functional memory model and a stochastic branch prediction model of our own. We set the parameters of the memory model (levels, capacity, hit/miss latency, etc.) to match those in gem5. The overall branch prediction accuracy in our stochastic model is set to match that in gem5 for each benchmark. This approach results in 9.5\% CPI error (geo-mean) for modeling the OoO core, which is almost twice the error with the higher-fidelity inputs (4.8\%) but possibly within an acceptable range depending on the use case, such as early-stage experiments when an accurate CAS may be unavailable but key design parameters may be generally known.

\textbf{Calipers Performance.}
In our experimental setup on a Skylake machine (with the Xeon Platinum 8160 CPU), we compare the speed of Calipers with gem5.
Measurements in Calipers include the time needed to read/parse the trace and build/analyze the DEG.
In a single-core execution, Calipers performed 54\% faster than gem5 on the InO configuration but had almost the same speed on the OoO configuration. If the approximate version of the OoO algorithm is used (Section~\ref{sec:alg}), Calipers performs 67\% faster than gem5 with a slight increase in CPI error (6.5\% rather than 4.8\%).

\begin{wrapfigure}{r}{0.40\linewidth}
\centering
\vspace{-10pt}
\includegraphics[width=0.85\linewidth]{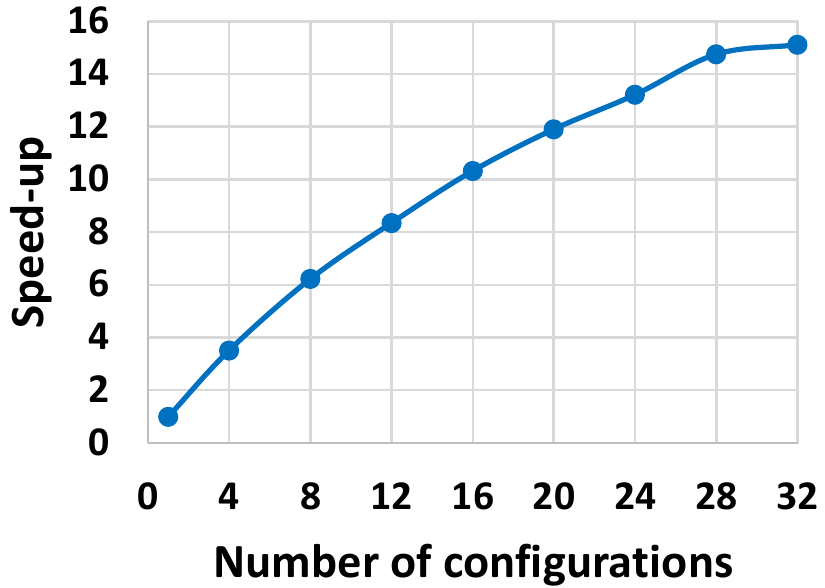}
\vspace{-5pt}
\caption{Speed-up of using vectorization over sequential modeling and analysis of different numbers of configurations.}
\label{fig:vector-speed}
\vspace{-5pt}
\end{wrapfigure}

With vectorized edge-weights (Section~\ref{sec:vec}), multiple configurations can be modeled in one, single-threaded execution.
Figure~\ref{fig:vector-speed} shows the speedup achieved by Calipers' vector-weighted DAG over a non-vectorized implementation in a single-threaded run.
The speedup, which is achieved by avoiding repeated graph construction, scales almost linearly with the vector size; up to 32 configurations can be easily modeled/analyzed concurrently at a speedup of $\sim$14$\times$.
Note that this is even without using SIMD instructions in our current implementation of edge-weight vectors; SIMD instructions may result in additional speedup.

\section{Applying Calipers to RISC-V}
\label{sec:risc5-model}

Architects often explore what-if scenarios to evaluate how much a core component is worth optimizing and how optimizations of different components interact with each other~\cite{fields:interaction-costs:micro:2003}. Answering such questions at an early design stage is crucial as it prevents unnecessary efforts and costs of over-optimizations. 
In this section, we do an in-depth bottleneck analysis of InO and OoO RISC-V cores, discuss two processor design scenarios and show how Calipers can provide insights.

\subsection{Bottleneck Analysis}
\begin{figure}[t]
\centering
\includegraphics[width=0.85\linewidth]{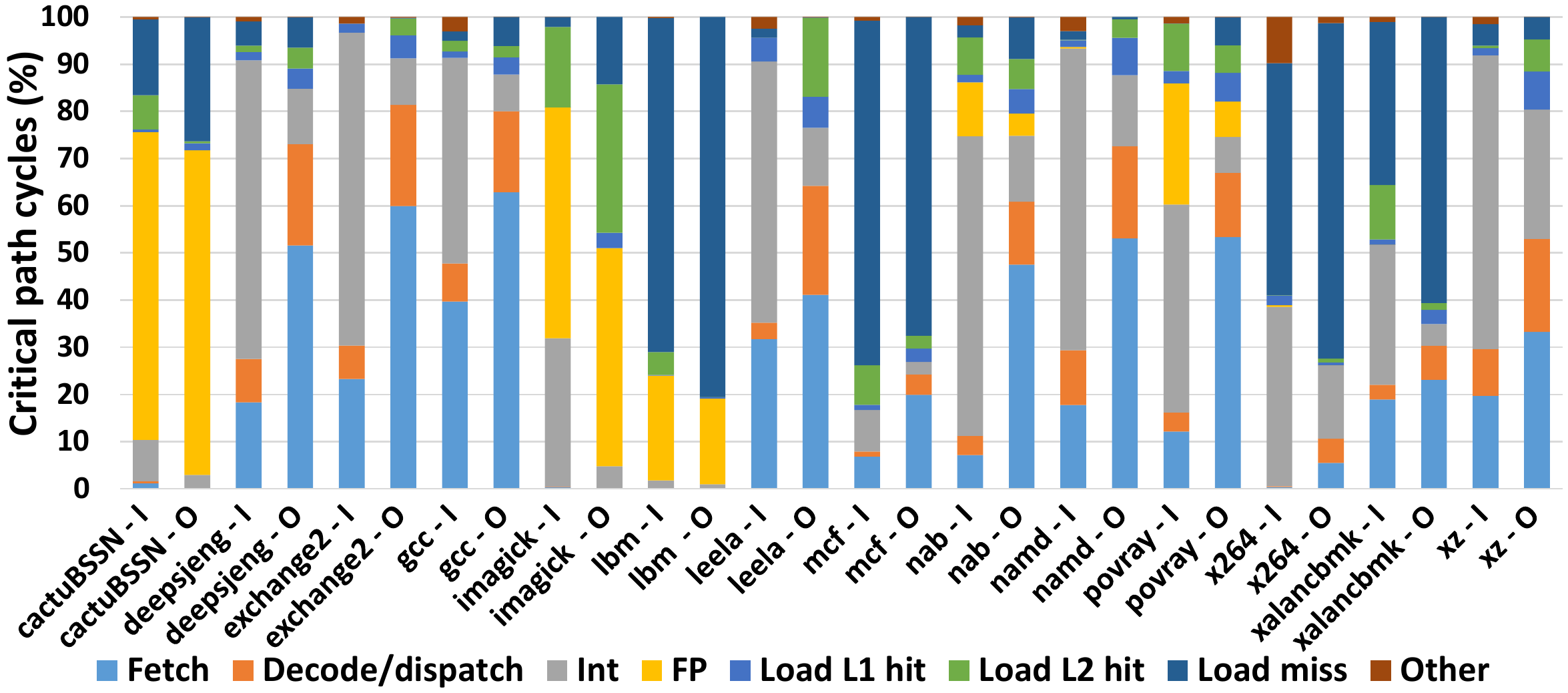}
\vspace{-10pt}
\caption{Breakdown of critical path cycles in the RISC-V in-order (I) and out-of-order (O) cores.}
\vspace{-15pt}
\label{fig:riscv-cp-cycles}
\end{figure}

Figure~\ref{fig:riscv-cp-cycles} shows the breakdown of critical path cycles in the InO and OoO RISC-V baseline configurations in Table~\ref{tab:gem5}. We highlight a few observations. There are several benchmarks where the front-end is a major bottleneck (\textit{deepsjeng}, \textit{exchange2}, \textit{gcc}, etc.) and which, as we will see next, benefit from optimizing fetch and branch prediction. In such benchmarks, fetch cycles have a higher share in the critical path of the OoO core in comparison to the InO core. The front-end bottleneck causes decode/dispatch cycles to appear in the critical path, which are, again, more conspicuous in the OoO core due to complexities like register renaming. Figure~\ref{fig:riscv-cp-cycles} also shows that Int cycles are more of a bottleneck in the InO core in contrast to the OoO core (e.g., in \textit{leela}, \textit{namd}, and \textit{xz}). This suggests that in the InO core, Int operations take longer or may be under-provisioned in the modeled system. This bottleneck breakdown, particularly for applications of interest, can guide architects on which structures, instructions, or operations should be optimized.

\subsection{What-if Scenario: Fetch and Branch Prediction}
\label{sec:what-if-fetch-bp}
\begin{figure}[t]
\centering
\includegraphics[width=0.70\textwidth]{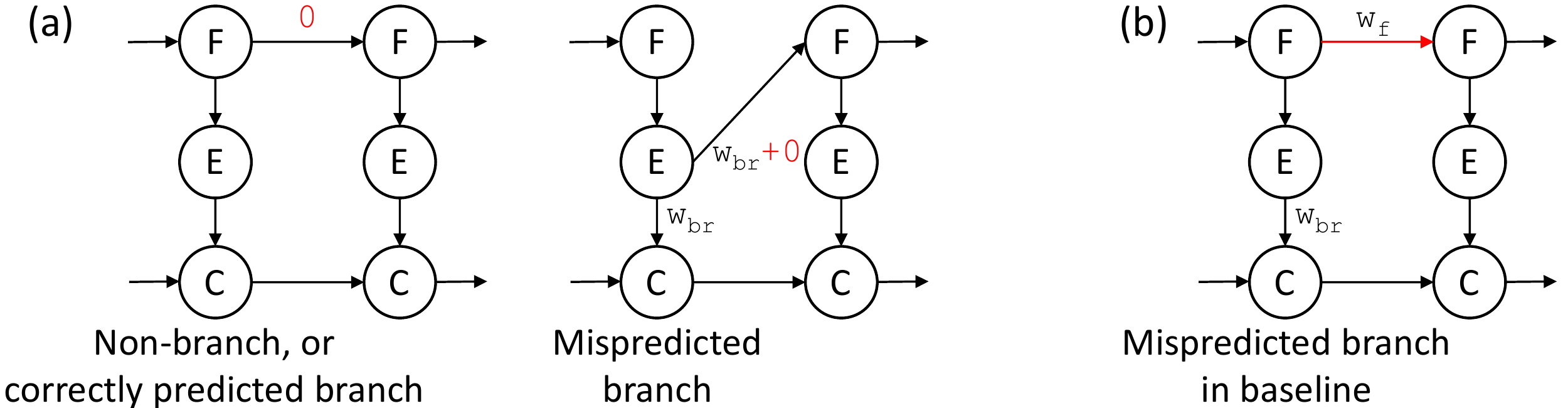}
\vspace{-5pt}
\caption{(a) Ideal fetch modeled by zeroing fetch cycles in edge weights. (b) Ideal branch prediction modeled by converting \edge{E$_n$}{F$_{n+1}$} (misprediction) to \edge{F$_n$}{F$_{n+1}$} (correct prediction).}
\vspace{-5pt}
\label{fig:ideal-fetch-bp}
\end{figure}

Next, we study a processor front-end design scenario.
High-performance processor implementation requires an effective front-end.
We study performance improvements from optimizing two key front-end components, the I-cache interface and the branch predictor, either in isolation or together. 
As such, we idealize these components to determine the opportunity. 
Calipers facilitates this study through simple modifications in the graph, as shown in Figure~\ref{fig:ideal-fetch-bp}. 
We model ideal fetch by discarding fetch cycles from the weight of fetch-related edges (Figure~\ref{fig:ideal-fetch-bp}(a)). 
Ideal branch prediction is modeled by transforming the edge \edge{E$_n$}{F$_{n+1}$} to \edge{F$_n$}{F$_{n+1}$}, in which instruction $n$ is a mispredicted branch in the baseline (Figure~\ref{fig:ideal-fetch-bp}(b)). 
Idealizing both fetch and branch prediction is achieved by combining both.
Note that idealizing only I-fetch in a simulator is not straightforward in case of unified I and D caches (e.g., L2), but can be easily done in Calipers.

\begin{figure*}[t]
\centering
\includegraphics[width=0.99\textwidth]{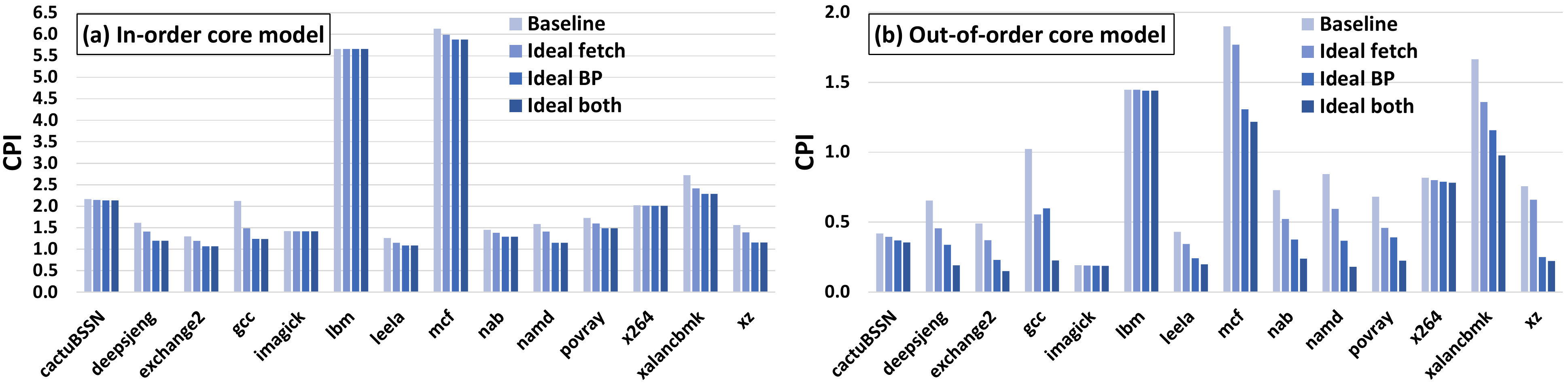}
\vspace{-5pt}
\caption{What-if analysis with Calipers for optimization candidates of a RISC-V core front-end.}
\vspace{-10pt}
\label{fig:risc5-opt-candidates}
\end{figure*}

Figure~\ref{fig:risc5-opt-candidates} illustrates the CPI improvement over the baseline configurations of RISC-V InO and OoO cores (Table~\ref{tab:gem5}).
We make two key observations. First, not all benchmarks benefit from idealizing fetch and branch prediction, e.g., \textit{x264}. 
In such benchmarks, their contribution to the critical path is small, and the core back-end is the bottleneck. 
This is supported by the critical path analysis in Figure~\ref{fig:riscv-cp-cycles}, which shows contributions of various components to a program's critical path.
The front-end is a very small contributor to the critical path of \textit{x264}
(see bars ``x264 - I/O'' in Figure~\ref{fig:riscv-cp-cycles}).

Second, in the benchmarks that benefit from these optimizations, e.g., \textit{gcc}, the behavior of InO and OoO cores is different. Figure~\ref{fig:risc5-opt-candidates}(a) shows that, in the case of the InO core, ideal branch prediction by itself brings benefits of ideal fetch and even more.
When branch prediction is idealized, the ``execution path'', i.e., the E$_{first} \rightarrow ... \rightarrow$E$_{last}$ path, becomes more critical than the ``fetch path'', i.e., the F$_{first} \rightarrow ... \rightarrow$F$_{last}$ path, because of the in-order issue constraint.  
In contrast, benefits of ideal fetch and ideal branch prediction complement each other in the OoO core (Figure~\ref{fig:risc5-opt-candidates}(b)), e.g., in \textit{gcc}.
As Figure~\ref{fig:riscv-cp-cycles} shows, the front-end bottleneck is more critical in the OoO core than the InO core in \textit{gcc} (see bars ``gcc - I/O'').
With ideal branch prediction, fetch cycles are still part of the critical path, and are reduced when fetch is optimized. 
Therefore, improving just the branch predictor in an InO core may suffice, but an OoO core can also benefit from optimizing instruction fetch.

\begin{wrapfigure}{r}{0.40\linewidth}
\vspace{-5pt}
\centering
\vspace{-5pt}
\includegraphics[width=0.40\textwidth]{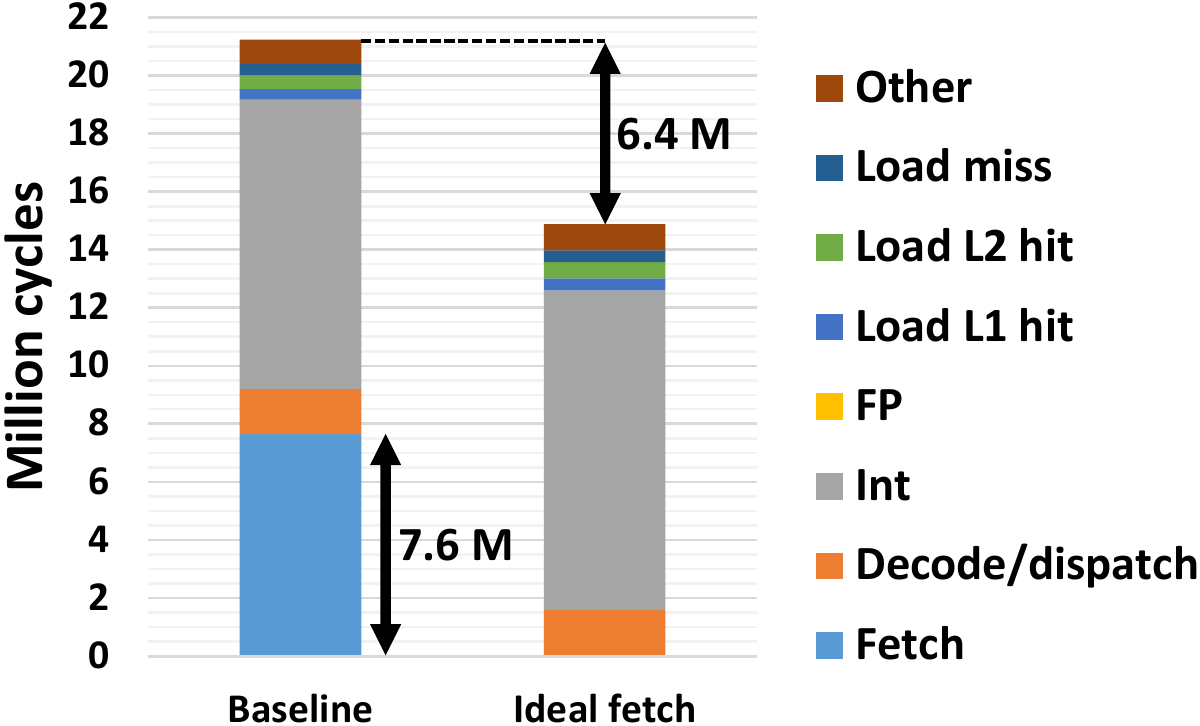}
\caption{Breakdown of critical path cycles of the \textit{gcc} benchmark in the InO core model.}
\vspace{-5pt}
\label{fig:gcc}
\end{wrapfigure}

We finally showcase how a stacked representation of spent cycles by itself may fall short of estimating the performance improvement from optimizing a candidate bottleneck. Figure~\ref{fig:gcc} shows the breakdown of the critical path cycles of \textit{gcc} in the InO core model. The breakdown implies that fully optimizing the front-end may at most save 7.6M cycles, but the exact improvement cannot be estimated because the secondary critical path is obscured in this breakdown. 
A CPI stack is similarly limited (in fact, even more so since it is an average metrics)~\cite{perf_cpi}. 
On the other hand, analysis of the ideal-fetch scenario through Calipers reveals the next critical path and the exact performance improvement, i.e., 6.4M cycles, as shown in the Figure.

\subsection{Criticality Insights Applied to Value Prediction}
\label{sec:value-pred}
As Calipers inherently computes critical instruction paths, users can easily use it to perform criticality-based experiments. We now investigate 
how adding criticality-awareness to value prediction can improve its effectiveness. We show findings similar to prior work, Focused Value Prediction (FVP)~\cite{bandishte2020focused}, and also expand the analysis by showing further potential in criticality-aware predictors.

Value prediction has been proposed to move ahead on execution even before the data has arrived from the memory.
Prior work has proposed adding criticality-awareness to reduce on-chip resources needed for prediction~\cite{bandishte2020focused}.
To show the benefit of criticality-awareness, we model two spectra of value prediction efforts in the OoO core of Table~\ref{tab:gem5}: (1) criticality-unaware that targets raw coverage, and (2) criticality-aware that prioritizes predicting critical instructions.
Following prior work~\cite{bandishte2020focused, sheikh2019efficient}, we investigate value prediction of only load instructions.

\begin{figure}
\centering
\includegraphics[width=0.7\linewidth]{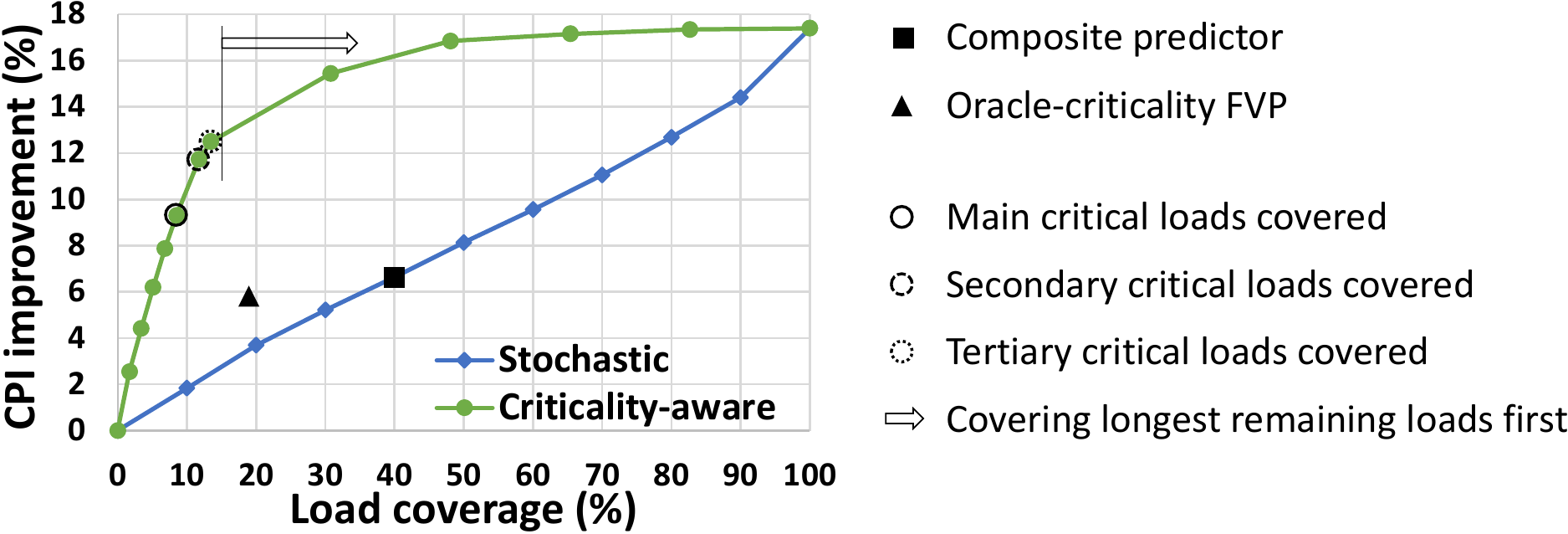}
\vspace{-5pt}
\caption{Value prediction performance of different approaches at different fractions of covered loads (coverage = fraction of value-predicted loads out of total loads).}
\label{fig:riscv-value-prediction}
\vspace{-10pt}
\end{figure}

We model {\it criticality-unaware} approaches by removing register-data-dependence edges sourcing from loads (0-100\% randomly, in 10\% increments) and re-calculating critical path to get CPI estimate (Figure~\ref{fig:riscv-value-prediction}). This represents predictors that have no regard for criticality---e.g., a criticality-unaware composite predictor~\cite{sheikh2019efficient}, which achieves 40\% coverage and 6.6\% speed-up, and lands straight on our stochastic line in Figure~\ref{fig:riscv-value-prediction}.

In contrast, we model {\it criticality-aware} approaches by removing only register-data-dependence edges sourcing from \textit{critical} loads (0--100\% randomly, in 20\% increments) and recalculating critical paths to get CPI estimates. We then do a second-pass removal (as new instructions become critical), third-pass removal, and finally, removal of remaining edges based on load latency (20\% increments). This represents predictors that target critical instructions first. 
The criticality-aware curve in Figure~\ref{fig:riscv-value-prediction} shows that most of the speedup can be realized by predicting $\sim$15\% of the loads. The data point of the criticality-aware FVP~\cite{bandishte2020focused}, which achieves 19\% coverage (including 40\% of main critical loads) and 5.8\% speedup, is also marked in Figure~\ref{fig:riscv-value-prediction}. 
Note that our critical-path calculations reveal yet more potential left on the table, e.g., similar performance is achievable at a lower, 5\% coverage.

These experiments show benefits of understanding and exploiting instruction-criticality, as well as showcasing flexibility of Calipers in performing related experimentation.

\section{Applying Calipers to EDGE}
\label{sec:edge-model}

In this section, we perform early-stage exploration of EDGE cores with Calipers. 
We study ISA- and microarchitecture-level bottlenecks and explore alternative EDGE block formats. 

We use a variant of the CLP TFlex processor~\cite{clp,robatmili:expoiting-criticality:hpca:2011}. 
It uses a 32-bit header to mark each block of up to 128 instructions. 
The header includes the block size in number of instructions. 
To achieve dense code, instructions can be variable in length, ranging from 2 to 13 bytes.
The design implements a 64-entry GPR, a centralized 128-instruction window, 8 Int-ALUs, 2 Int-MULs, 2 Int-DIVs, 2 FP-ALUs, 2 FP-MULs, 1 FP-DIV, 2 load-store units, and a 32-entry load-store queue. 
It can speculatively execute up to 4 blocks concurrently if each has no more than 32 instructions.

Since we seek to do a high-level analysis of two major EDGE architecture features for which precise cost models are not essential, we use a statistical model for various costs.
We assume branch prediction accuracy of 95\%, 98\% I-cache hit rate, 90\% D-cache hit rate, 2-cycle cache hit latency, 2 store commits per cycle and 4 GPR commits per cycle. 
We assume a slightly optimistic cache miss latency of 10 cycles so that loads and stores do not dominate and obscure other events in the critical path. The instruction window is partitioned into four segments of 32 instructions each, with a cycle cost of communicating results to a different partition. 
We use benchmarks from the SPEC CPU 2006 suite, with reference inputs, and do experiments on 10 million instruction blocks after the first 100 million blocks of each benchmark.
We generate instruction traces using a functional simulator.

\begin{figure*}[t]
\centering
\subfloat[Instructions]{\includegraphics[width=0.5\textwidth]{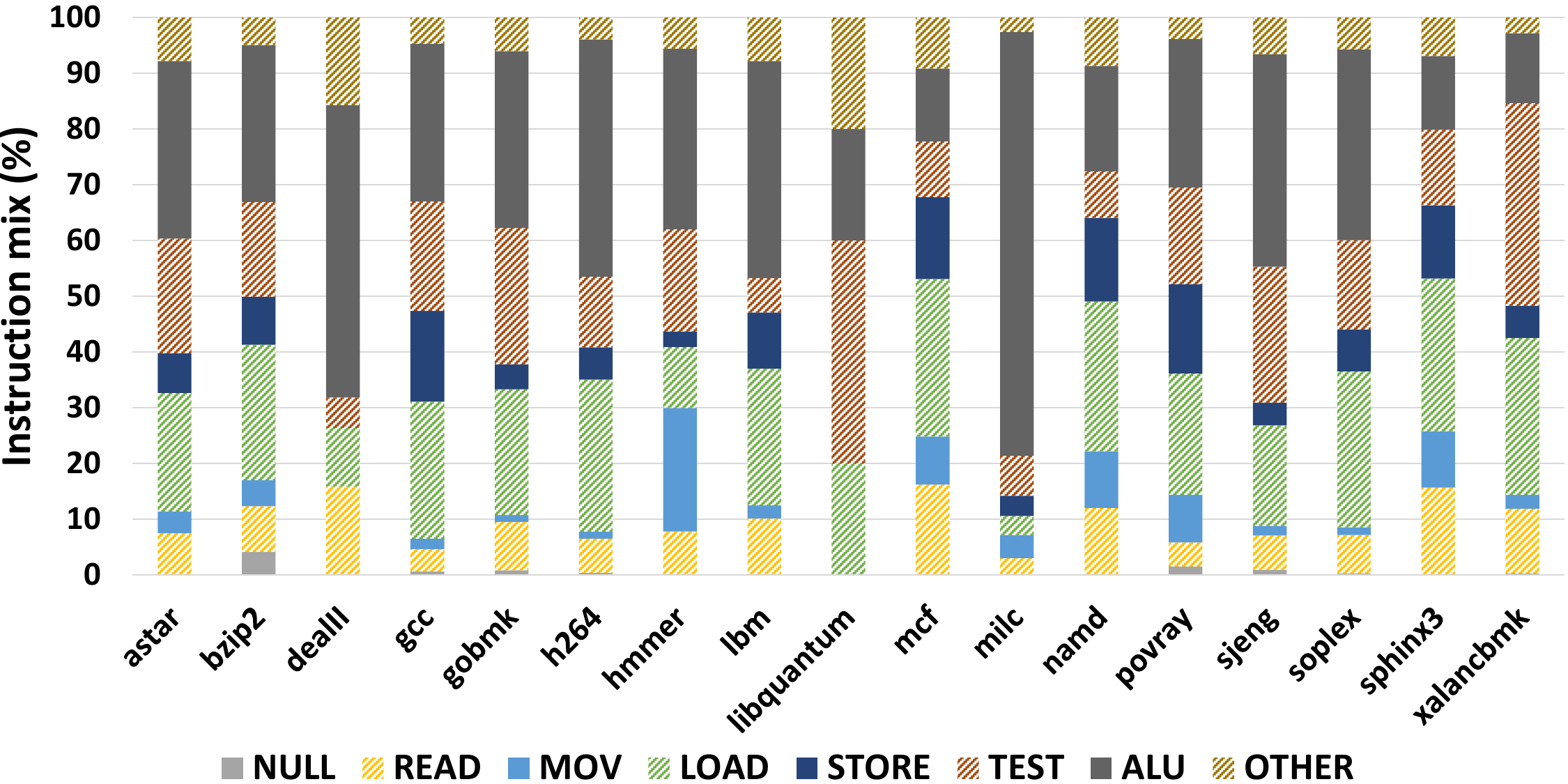}\label{fig:critical-path-instructions}}
\subfloat[Cycles]{\includegraphics[width=0.5\textwidth]{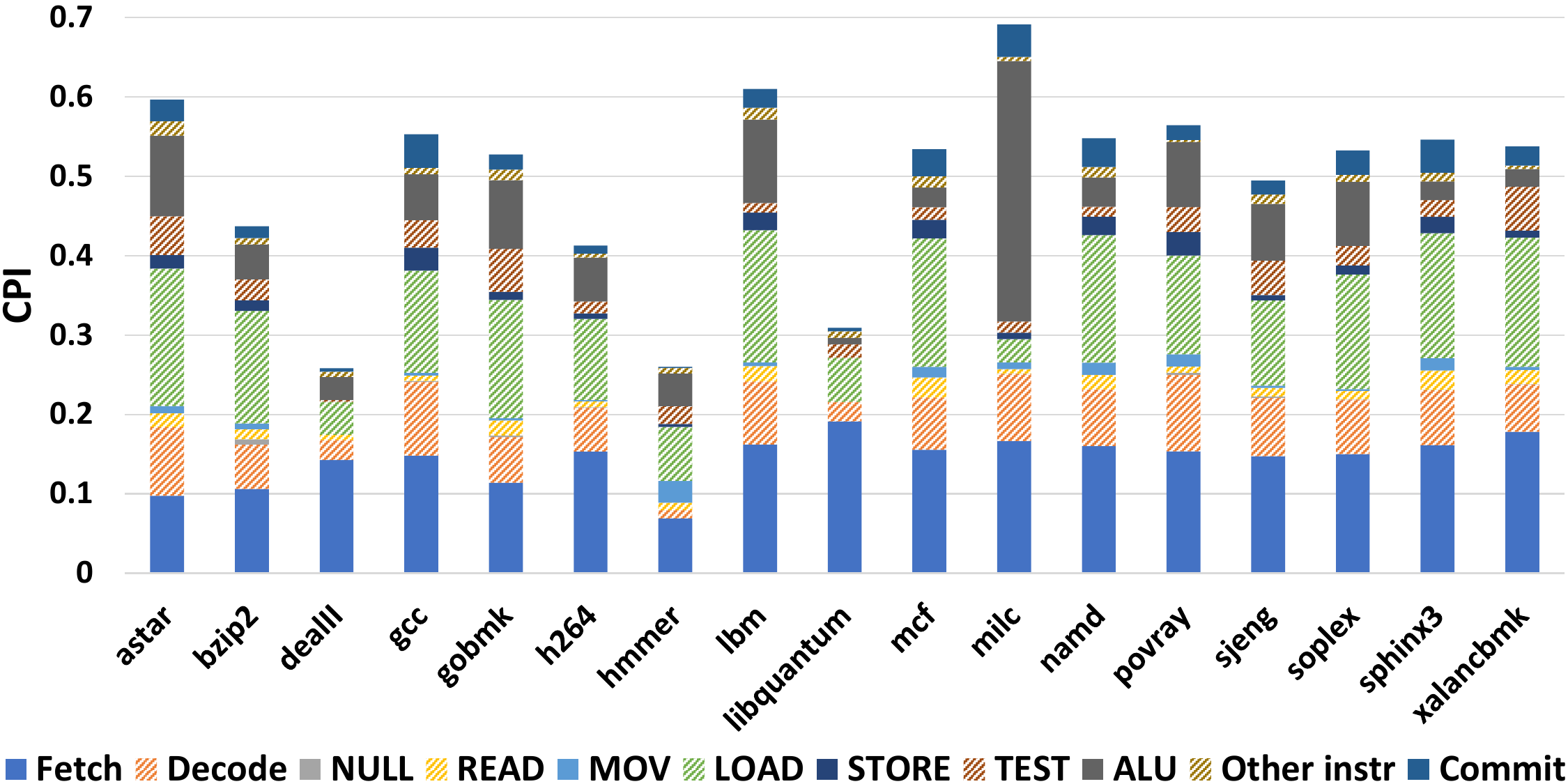}\label{fig:critical-path-cycles}}
\vspace{-10pt}
\caption{Breakdown of (a) committed instructions and (b) cycles in critical path of EDGE core with 4-segment block format.}
\vspace{-15pt}
\label{fig:critical-path-breakdown}
\end{figure*}

\subsection{Bottleneck Analysis}
\label{sec:edge-bottlenecks}
We apply Calipers to analyze bottlenecks in the EDGE core, with a focus on EDGE-specific instructions, \texttt{READ}, \texttt{MOV}, and \texttt{NULL}.
Figure~\ref{fig:critical-path-breakdown} depicts the instruction mix and breakdown of cycles in the critical path of different benchmarks. In Figure~\ref{fig:critical-path-breakdown}(a), we observe that other than conventional compute and load/store instructions, EDGE-specific data movement instructions, i.e., \texttt{MOV} and especially \texttt{READ}, make up a considerable part of the critical path (13.5\% of instructions, on average). However, \texttt{NULL} instructions---another EDGE-specific artifact---have a tiny share in the critical path (0.5\% of instructions, on average).

The breakdown of cycles in Figure~\ref{fig:critical-path-breakdown}(b) shows that even a 5\% rate of branch misprediction causes \textit{Fetch} and \textit{Decode} cycles to make up a significant part of the critical path. A subset of these cycles correspond to \texttt{READ} and \texttt{MOV} instructions, which form a considerable share in the critical path as seen above. Fetching, decoding, and executing \texttt{READ}s and \texttt{MOV}s together take 9.9\% of critical path cycles in this experiment, on average. Such large costs of \texttt{READ}s and \texttt{MOV}s indicate that EDGE architectures must find ways to minimize their impact.

\subsection{What-if Scenario: Block Format}
\label{sec:edge-blocks}
Next, we investigate alternate organization of bytes in EDGE instruction blocks. 
The variable-length instruction encoding can complicate the hardware decode logic in a multi-issue design, possibly impacting achievable clock frequency. 
The EDGE block structure presents an opportunity to organize instructions within a block to simplify the decode logic. However, care must be taken to not impact the issue rate.

\begin{figure}[t]
\centering
\subfloat{\includegraphics[width=0.45\textwidth]{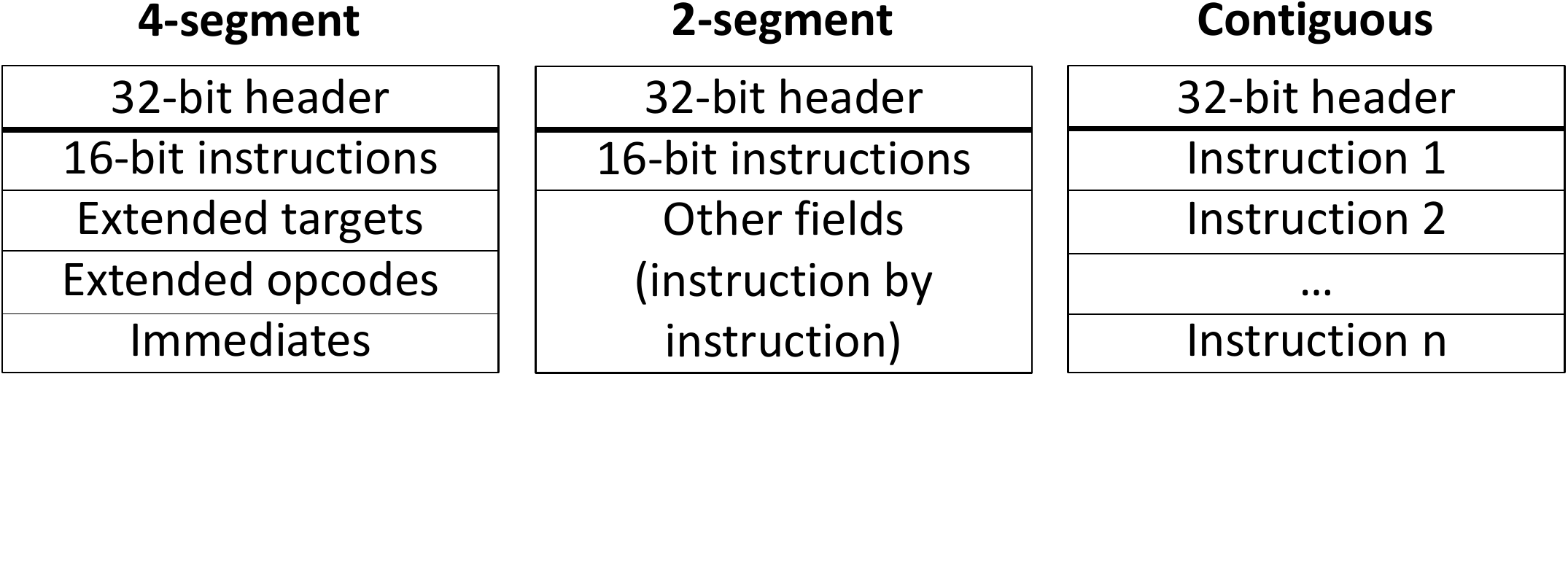}}
\subfloat{\includegraphics[width=0.55\textwidth]{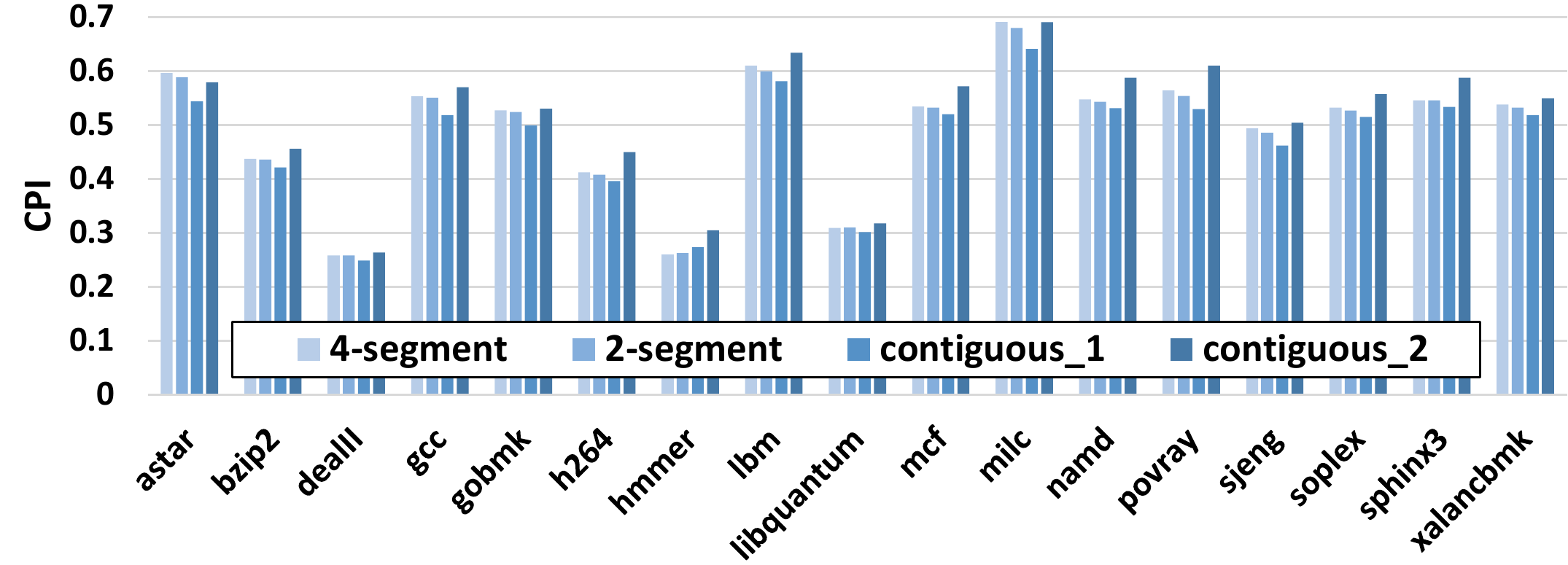}}
\vspace{-12pt}
\caption{Different block formats for EDGE and estimated CPI.}
\vspace{-17pt}
\label{fig:block-format}
\end{figure}

Each EDGE instruction comprises a 16-bit base and additional optional bytes. The base specifies the primary opcode and is used to identify the length of the instruction and the additional fields.
The optional additional fields include extended targets, an extended opcode, and/or an immediate value. 
Figure~\ref{fig:block-format}(a) shows three block formats that we study: 4-segment, 2-segment, and contiguous.
The 4- and 2-segment formats simplify instruction decode by packing the fixed-size base of block instructions in one segment.
By grouping purpose-specific fields together in the 4-segment format, the unpacking of various fields in an instruction is further simplified.
The contiguous format, wherein instructions are organized linearly, complicates unpacking and decoding because of unaligned fields in the instruction-fetch stream.
However, it makes earlier instructions in the block available for execution sooner in comparison to the other formats.
Therefore, a tradeoff exists between decoder complexity, which may affect clock frequency, and scheduling delays.

We can quantify the effect of different block formats on performance by adjusting the weight of the \edge{BF}{E} edges (Figure~\ref{fig:block-format}(b)).
There is little difference between the CPI of the 4- and 2-segment formats.
As a result, based on the implementation complexity, a choice can be made between the two.
On the other hand, assuming no impact on clock frequency, the contiguous format (contiguous\_1 bars in the Figure) results in 4.6\% CPI improvement, on average, over the 4-segment organization (up to 8.9\%) in all benchmarks except \textit{hmmer} (5.2\% worse).
In the contiguous format, sooner availability of earlier block instructions improves performance in most benchmarks. 
Interestingly, the CPI of the contiguous format is lower than the 4-segment format for \textit{hmmer}. The reason, at a high level, is that the optional fields of the earlier block instructions delay availability of the following instructions, which happen to be on the critical path.

We also analyzed a microarchitectural choice that adds an extra cycle in the decode pipeline assuming that the contiguous format impacts the clock frequency. The extra cycle increases the CPI by 9.0\% over the no-penalty performance and by 4.6\% over the 4-segment format (contiguous\_2 bars in Figure~\ref{fig:block-format}(b)).
Thus, designers may prefer to optimize the decoder circuit design in order to take advantage of the contiguous block format.

Note that evaluating this feature using a simulator would have required compiler changes besides modifying the simulator, but was not when using Calipers.

\vspace{-3pt}
\section{Related Work}
\vspace{-3pt}
\label{sec:related}

\textbf{Graph-based Approaches.}
Fields et al. presented a model of dependence graphs to profile critical paths through a trace of instruction execution~\cite{10.1145/379240.379253,1003561}. They used this model to dynamically predict critical instructions in the hardware and steer or schedule them to improve performance.
This model was subsequently extended to build a framework to analyze microarchitectural bottlenecks and draw insights in effectiveness of example design choices~\cite{fields:interaction-costs:micro:2003}.
Others have proposed enhancements to the methodology by using profile information and random traces from the program code~\cite{10.1145/1356058.1356068}.
Tanimoto et al. enhanced the model to match gem5's accuracy~\cite{enhanced-graph}. Li et al. extended the dependence graph model to study criticality in shared memory multiprocessors.
Saidi et al. applied a similar approach to analyze bottlenecks at a system level~\cite{10.1145/1555754.1555800}.
However, unlike Calipers, these prior proposals do not manipulate the graph to explore ISA or microarchitecture techniques, and hence do not model dynamic resource conflicts and scheduling, nor perform vectorized analysis. Calipers also allows more types of inputs besides CAS and hardware, such as from functional simulators and statistical models, to permit accuracy vs. speed tradeoffs.
A related work presented preliminary what-if analysis for EDGE using graph manipulation for few benchmarks~\cite{cal}.

Although Nowatzki et al.~\cite{tdg_techreport} also model program execution using dependence graphs and explore accelerator designs by modifying the graphs, they do not describe scheduling of instructions in pipelined OoO superscalar processors and modeling structural hazards. 
In contrast, we discuss these issues and related algorithms in depth. 
Modeling them accurately is important since resource availability and requirements, and hence scheduling, could change with alternative designs. 
Additionally, we model different types of ISAs, show breakdowns of critical paths, and propose vectorized analysis for rapid exploration. 

A number of related works have used dependence graphs to enhance processor~\cite{1540948,robatmili:expoiting-criticality:hpca:2011} and cache hierarchy designs~\cite{8416821}, improve code generation~\cite{nagarajan:critical-path-trips:ispass:2006, 10.1145/1366230.1366267}, modify run time scheduling algorithms~\cite{robatmili:expoiting-criticality:hpca:2011}, evaluate DVFS schemes~\cite{10.1145/1366230.1366267, 10.1145/2016604.2016654}, and design accelerators~\cite{nowatzki:exploring-potential-heterogeneous-models:isca:2015,tdg_techreport}. Calipers is a more generic dependence-graph tool and can be applied to similar efforts for offline analysis to explore techniques and to inform  heuristics for run time scheduling during program execution.

\textbf{Non-graph Approaches.}
Non-graph approaches have also been proposed to identify microarchitecture bottlenecks~\cite{topdown_perf,perf_cpi,multistage_cpi}.
Mechanistic models~\cite{ino_mechanistic, highlevel_mechanistic, ooo_mechanistic, interval} view program execution as a sequence of intervals, each consisting of a base part of efficient execution followed by a stalled part due to hazards such as branch mispredictions, long-latency cache misses, etc. They estimate performance penalties using hand-constructed analytical models, parameterized by microarchitectural characteristics and event costs such as miss latencies, and the dispatch rate by applying Little's Law on the critical path length in instruction windows. They model structural hazards by considering patterns of instruction sequences~\cite{ino_mechanistic} and the number of access contentions for ports in instruction windows~\cite{highlevel_mechanistic}. Interval simulation~\cite{interval, sniper} uses the interval model for multi-core timing simulation.

While both the interval approach and Calipers raise the level of abstraction compared to hardware execution and CAS, interval modeling directly models the performance penalties, or effects, on program execution whereas Calipers models the constraints imposed by microarchitecture and program characteristics and lets the program effects be computed by graph traversal algorithms. In interval modeling, various penalty scenarios are enumerated and analytical models are carefully constructed for each scenario taking into account assumptions about how different stall conditions interact. This can lead to missed scenarios~\cite{topdown_perf} and difficulty in tracking which models should be updated, and correctness of those updates, for what-if scenarios involving changes to microarchitectural characteristics. In contrast, Calipers enumerates edge types to model various constraints and does not limit the extent of interactions and overlaps of stall conditions. As we showed, a wide spectrum of designs and what-if scenarios, including ISA modifications, can be modeled by making simple and easy-to-follow changes to the graph structure. Further, we also employed vectorized analysis, and investigated structural hazard modeling in depth, including comparing the two modeling approaches.
 
GDP~\cite{GDP} combines mechanistic models with a dependence graph of load requests that miss in L1 cache to analyze performance and develop a new LLC management policy. Combining the mechanistic model with full-program dependence graph analysis using Calipers could be interesting future work.

\vspace{-5pt}
\section{Conclusion}
\label{sec:conclusion}

We proposed Calipers, a framework that models program execution on complex architectures using vector-weighted dynamic event-dependence graphs. 
Calipers facilitates different types of complex performance analyses on ISA features and microarchitectural techniques by abstracting data, control, and resource dependencies through weighted edges in the graph. 
Calipers can use various cost models and enable rapid design space exploration.
Algorithms to model instruction scheduling in in-order and out-of-order cores were presented.
We applied Calipers to perform insightful analyses on in-order and out-of-order RISC-V and EDGE cores and answer real-life what-if questions.
Calipers can be a valuable tool in the architect's toolkit.

Calipers provides wide latitude in what users can explore.
The user can change just the edge weights and/or alter the graph to model ideas ranging from ISA features, to compiler techniques, to microarchitecture techniques. 
Like with any other tool, it must be used judiciously.
Although new features can be modeled by introducing new types of edges and vertices, ultimately, users must ensure that changes to the graph do not compromise its integrity with respect to the inputs used to construct the original graph.

In future work, we plan to explore graph pattern matching to automatically detect acceleration opportunities. Moreover, modeling multiple cores/threads by applying appropriate costs and orders for shared memory accesses is another aspect for future work.


\bibliographystyle{ACM-Reference-Format}
\bibliography{refs}

\end{document}